\documentstyle[11pt,aaspp4]{article}
\begin{document}

\title{The DIRECT Project: Influence of Blending on the Cepheid
Distance Scale. II.~Cepheids in M33}

\author{B. J. Mochejska\altaffilmark{1}}
\affil{Copernicus Astronomical Center, 00-716 Warszawa, Bartycka 18}
\affil{\tt e-mail: mochejsk@camk.edu.pl}
\author{L. M. Macri, D. D. Sasselov\altaffilmark{2}, 
K. Z. Stanek}
\affil{Harvard-Smithsonian Center for Astrophysics, 60 Garden St.,
Cambridge, MA~02138}
\affil{\tt e-mail: lmacri, dsasselov, kstanek@cfa.harvard.edu}
\altaffiltext{1}{Visiting Student, Harvard-Smithsonian Center 
for Astrophysics}
\altaffiltext{2}{Alfred P. Sloan Foundation Fellow}

\begin{abstract}
We investigate the influence of blending on the Cepheid distance
scale. Blending is the close association of a Cepheid with one or more
intrinsically luminous stars. High-resolution {\em HST} images are
compared to our ground-based data, obtained as part of the DIRECT
project, for a sample of 102 Cepheids in the M33 galaxy. The average
(median) flux contribution from luminous companions not resolved on
the ground-based images in the $B$, $V$ and $I$ bands ($S_B$, $S_V$,
$S_I$) is about 29\% (15\%), 24\% (14\%), 30\% (21\%) of the flux of
the Cepheid. For 64 Cepheids with periods in excess of 10 days the
average (median) values of $S_B$, $S_V$, $S_I$ are 20\% (10\%), 16\%
(7\%), 23\% (14\%). This shows that, depending on the sample of
objects chosen, the distance derived from our ground-based $V$-band
magnitudes for the M33 Cepheids would be systematically underestimated
by about 8\%-11\% (3\%-6\%).

Using artificial star tests we study crowding and blending as separate
phenomena. These tests indicate that the effect of blending could be
more significant than crowding in regions of lower surface brightness,
below 21.6 mag/\sq\arcsec. We also find indications that in the case
of long period Cepheids ($P>10^d$) companions from blending are
on average bluer than companions introduced by crowding.

Our ground-based resolution in M33 corresponds to the {\em HST}
resolution at about $11$ Mpc. We estimate the effect of blending at
resolutions corresponding to a range of distances as observed with the
{\it HST} using the {\it HST} M33 data as the template unblended
population. We compare our results to those obtained by other groups.
\end{abstract}

\section{Introduction}

As the number of extragalactic Cepheids discovered with {\em HST}
continues to increase and the value of $H_0$ is sought from distances
based on these variables (e.g. Saha et al.~1999; Freedman et
al.~2001), it becomes even more important to understand various
possible systematic errors which could affect the extragalactic
distance scale. Currently, the most important systematic is a bias in
the distance to the Large Magellanic Cloud, which provides the
zero-point calibration for the Cepheid distance scale. The LMC
distance is very likely significantly shorter than usually assumed
(e.g. Udalski 2000; Stanek et al.~2000, Fitzpatrick et al.~2001).
It still might be considered uncertain at the $\sim 10$\% percent
level (e.g. Jha et al.~1999). Another possible systematic, the
metallicity dependence of the Cepheid Period-Luminosity (PL) relation,
is also very much an open issue, with empirical determinations ranging
from 0 to $-0.4$ mag dex$^{-1}$ (Freedman \& Madore 1990; Sasselov et
al.~1997; Kochanek ~1997; Kennicutt et al.~1998).

In this paper we investigate the influence of blended stellar images
on the derived Cepheid distances. We define {\em blending} as the
close projected association of a Cepheid with one or more
intrinsically luminous stars, which cannot be detected within the
observed point-spread function (PSF) by photometric analysis (e.g.,
DAOPHOT, DoPHOT). Such blended stars are mostly other young stars
which are physically associated $-$ from actual binary and multiple
systems to companions which are not gravitationally bound to the
Cepheid. Blending is thus a phenomenon different from {\em crowding}
or {\em confusion noise}; the latter occurs in stellar fields with a
crowded and complex background due to the random superposition of
stars with different luminosities. In this paper, we are concerned
with blending due to wide systems.

The issue of the effect of blending on the derived Cepheid distances
has received some attention since the appearance of our first paper on
M31 Cepheids (Mochejska et al. 2000, hereafter Paper I). An estimate
of the magnitude of this effect for remote galaxies was obtained by
Stanek \& Udalski (1999) using the LMC stellar population as the
template. Their analysis indicated that blending could be a
significant (up to 20\%) source of systematic error in distances
determined using Cepheids. Saha, Labhardt \& Prosser (2000) studied
the effect of confusion noise on Cepheid derived distances using
artificial star tests. Their experiment showed that this effect is
within 0.1 mag for NGC 4639 at a distance of 25 Mpc. A further study
of the effects of crowding and confusion noise was conducted by
Ferrarese et al. (2000) as part of the HST Key Project on the
Extragalactic Distance Scale, employing artificial star tests. They
estimated that the photometry of Cepheids could be biased too bright
by up to 0.2 mag due to crowding. Under the assumption that multiple
epoch data can be used to reject most of the affected Cepheids, they
derived an upper limit of 0.02 mag for the bias in the distance scale
due to crowding. Gibson, Maloney \& Sakai (2000) presented three
empirical tests of blending. Their results were inconclusive and only
indicated at a 1 $\sigma$ level that blending for Cepheids located in
the LMC bar is not representative for the distant galaxies observed by
the HST Key Project. An uncertainty of $^{+5}_{-0}\%$ due to crowding
and/or blending was included into the systematic error budget in the
HST Key Project final results paper (Freedman et al. 2001). Macri et
al. (2001a) studied the effect of blending in an inner field of M101
by scaling to its distance the stellar populations observed with
NICMOS in M31 and M81. They concluded that this effect could be
significant and lead to a 0-0.2 mag bias in the derived distance
modulus.

We investigate the effects of stellar blending on the Cepheid distance
scale by studying two Local Group spiral galaxies, M31 and M33. The
results for M31 were presented in Paper I. In this paper, second of
the series, we concentrate on M33, which for the purposes of this
paper we assume to be located at a distance of 850 kpc. As part of the
DIRECT project (Kaluzny et al. 1998), we have collected an extensive
data set for this galaxy, and thus far have discovered $\sim 400$
Cepheids, among other variables. We identify some of these Cepheids on
archival {\em HST}-WFPC2 images and compare them to our ground-based
data to estimate the impact of blending on our photometry, taking
advantage of their superior resolution -- the $0\farcs12$ FWHM on the
WFPC2 camera (Ferrarese et al. 2000) corresponds to $\sim 0.5$ pc at
the distance of M33, compared to $\sim 6$ pc for the ground-based
data.

This paper is organized as follows: Section 2 describes the
ground-based and {\em HST} data and the applied reduction procedures.
In Section 3 we discuss the process of identifying Cepheids on {\em
HST} WFPC2 images. In Section 4 we present the Cepheid blending
catalog and discuss it in Section 5. In Section 6 we present the
results of artificial star tests. In Section 7 we show an example of
an extremely blended Cepheid. The implications of our results for
remote galaxies are studied in Section 8. In Section 9 we discuss the
results of other investigations of Cepheid blending and/or crowding.
The concluding remarks are to be found in Section 10.

\section{Observations and Data Reduction}

\subsection{Ground-based Data}

The ground-based data were obtained as part of the DIRECT project
between September 1996 and October 1997 during 42 nights on the
F. L. Whipple Observatory 1.2~m telescope and 10 nights on the
Michigan-Dartmouth-MIT 1.3~m telescope. Three $11'\times11'$ fields A,
B and C with a scale of 0.32 \arcsec/pixel were monitored, located
north, south and southwest of the center of M33, respectively.  The
data for fields A and B has been reduced and the $BVI$ photometry of
251 Cepheid variables published by Macri et al. (2001b). The reduction
of the field C data, which is now part of a larger field Y, is in
progress and the results will be reported by Stanek et al. (2001). The
applied reduction, calibration and variable selection procedures were
described by Kaluzny et al. (1998).

\subsection{HST data}

The archival {\em HST}-WFPC2 data used in this paper were retrieved
from the Hubble Data Archive. We selected images overlapping our M33
fields observed from the ground, taken in filters F439W, F450W
(roughly $B$), F555W, F606W ($\sim V$) and F814W ($\sim I$). The pixel
scales of the Wide Field (WF) and Planetary Camera (PC) chips are
0.1 and 0.046 \arcsec/pixel, respectively.

The HST data overlapping our fields A and B had already passed through
the standard preliminary processing and calibration procedures prior
to its placement in the Archive. The HST data for field C, retrieved
later, was calibrated by the On-The-Fly Calibration (OTFC) system at
the time when our request was processed. The standard pipeline
calibration and the OTFC are fully described in the {\em HST} Data
Handbook.

The images were corrected under IRAF\footnote{IRAF is distributed by
the National Optical Astronomy Observatories, which are operated by
the Association of Universities for Research in Astronomy, Inc., under
cooperative agreement with the NSF.} for the geometric distortion in
the WFPC2 optics and for bad pixels (Stetson 1998). Whenever possible,
the images were combined to remove cosmic rays. The photometry was
extracted using the DAOPHOT/ALLSTAR package (Stetson 1987, 1992). A
more detailed description of the reduction procedure can be found in
Paper I.

It should be noted that the {\em HST} photometry has not been
calibrated to any standard system and therefore instrumental
magnitudes are used throughout this paper. This has, however, no
bearing on the results presented in this paper, since they are
strictly based on differential photometry.

\section{The Identification of DIRECT Cepheids in {\em HST} Data}

The preliminary identification of DIRECT Cepheids on the {\em HST}
frames was performed under SAOimage ds9\footnote{SAOimage ds9 was
developed under a grant from NASA's Applied Information System
Research Program (NAG5-3996), with support from the Chandra Science
Center (NAS8-39073)} by a visual comparison of the {\em HST} data
matched via World Coordinate System (WCS) information to our
ground-based $V$-band templates. The WCS information for the
ground-based template image header was derived by matching stars from
the image to the USNO A2.0 Catalog stars (Monet et al.~1996) using the
WCSTools image astrometry toolkit (Mink 1999).

Our list of Cepheids consists of an early version of the Macri et
al. (2001b) catalog supplemented by field C Cepheids with periods in
excess of 10 days. We have matched a total of 102 Cepheids to {\em
HST} data: 48 in field A, 39 in field B and 15 in field C. There are
64 Cepheids with periods over 10 days in this sample.

\begin{figure}[!t]
\plotfiddle{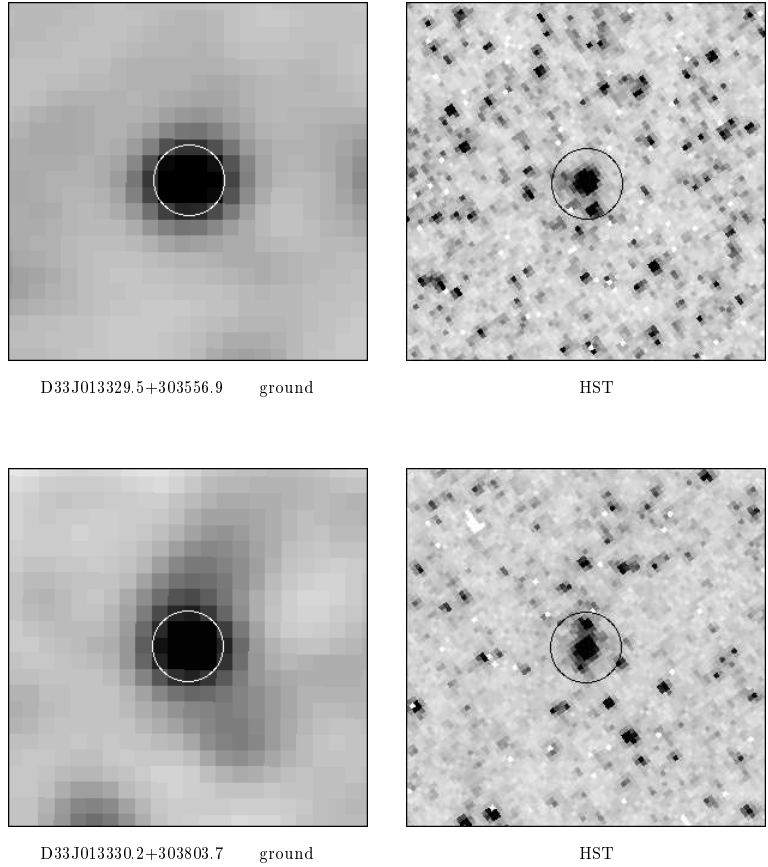}{10.5cm}{0}{60}{60}{-185}{-120}
\caption{A comparison of the ground-based and {\em HST} images of two
Cepheids: D33J013329.5+303556.9 and D33J013330.2+303803.7. Circles
$0\farcs75$ in radius are drawn centered on the Cepheids. }
\label{fig:cfig}
\end{figure}

In some cases a Cepheid which appeared to be a single star on the
$V$-band template was resolved into multiple stars on an {\em HST}
image. Two examples are shown in Figure \ref{fig:cfig}. The images
plotted in the left panels are from the $V$-band template image (FWHM
$\sim 1\farcs4$) and those in the right panel were taken with the WF
chips of the WFPC2 camera. Circles $0\farcs75$ in radius are drawn
centered on the Cepheids. The Cepheid D33J013329.5+303556.9 is shown
in the upper panels. On the ground-based template it appears as a
single star, while on the {\em HST} image it is resolved into two
objects differing by 2.5 magnitudes in brightness. The lower panels
show the case of D33J013330.2+303803.7, where the Cepheid, single on
the ground $V$-band template, is resolved by {\em HST} into three
stars, with the two companions fainter by 2 and 3 magnitudes.

\begin{figure}[t]
\plotfiddle{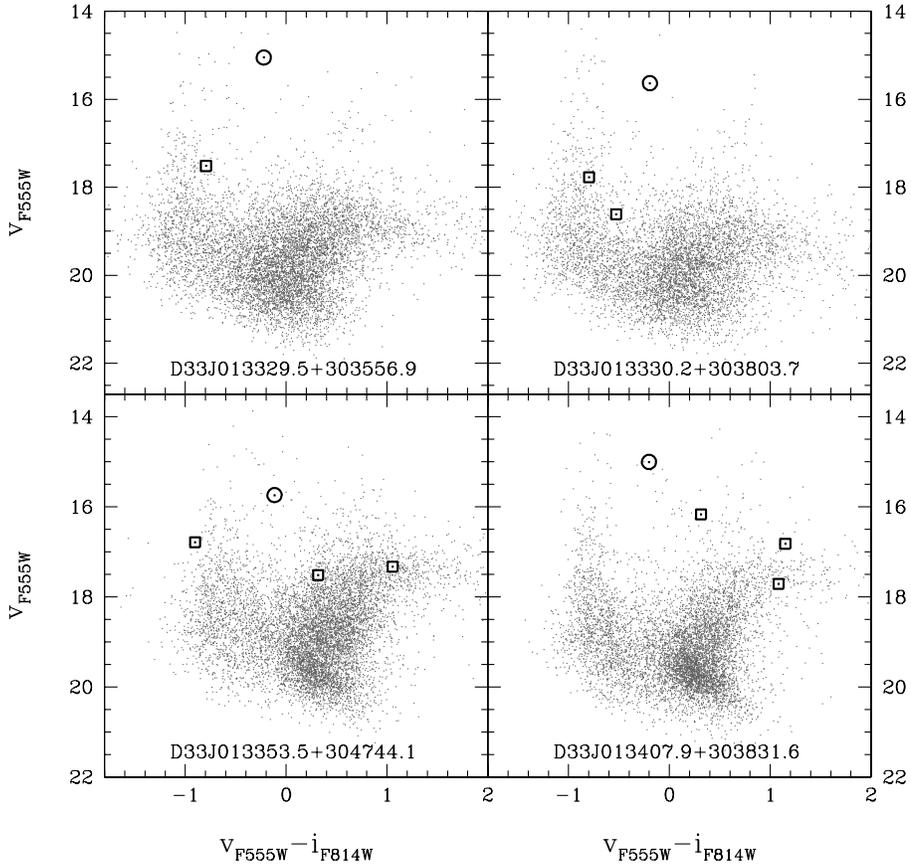}{10cm}{0}{60}{60}{-195}{-97}
\caption{Selected ($v_{F555W}, v_{F555W}-i_{F814W}$) instrumental
color-magnitude diagrams for Cepheids and their companions within
$0\farcs75$ based on {\em HST} data. The Cepheids are denoted by
circles and their companions by squares. Stars from the same image are
plotted in the background.}
\label{fig:cmd}
\end{figure}

To help confirm the Cepheid nature of the objects selected,
instrumental color-magnitude diagrams (CMDs) were constructed from
{\em HST} data, whenever photometry in two bands was available. A few
representative ($v_{F555W}, v_{F555W}-i_{F814W}$) CMDs are shown in
Figure~\ref{fig:cmd}. The Cepheids are denoted by circles and their
companions by squares. Stars from the same image are plotted in the
background for reference. The upper panels present two Cepheids with
blue blends (those shown in Fig. \ref{fig:cfig}). D33J013329.5+303556.9, 
in the upper left panel, has a blue companion contributing 10\% of its
flux in $V$ and 6\% in $I$. D33J013330.2+303803.7, in the upper right
panel, has two blue blends at the level of 14\% and 6\% in $V$ and 8\%
and 5\% in $I$. In the lower left panel we show the Cepheid
D33J013353.5+304744.1 with one blue and two red companions at the
level of 33\%, 24\%, 20\% in $V$ and 18\%, 64\%, 29\% in $I$.
D33J013407.9+303831.6 in the lower right panel is a case with three
red blends at the level of 34\%, 19\%, 8\% in $V$ and 55\%, 10\%, 27\%
in $I$.

\begin{figure}[t]
\plottwo{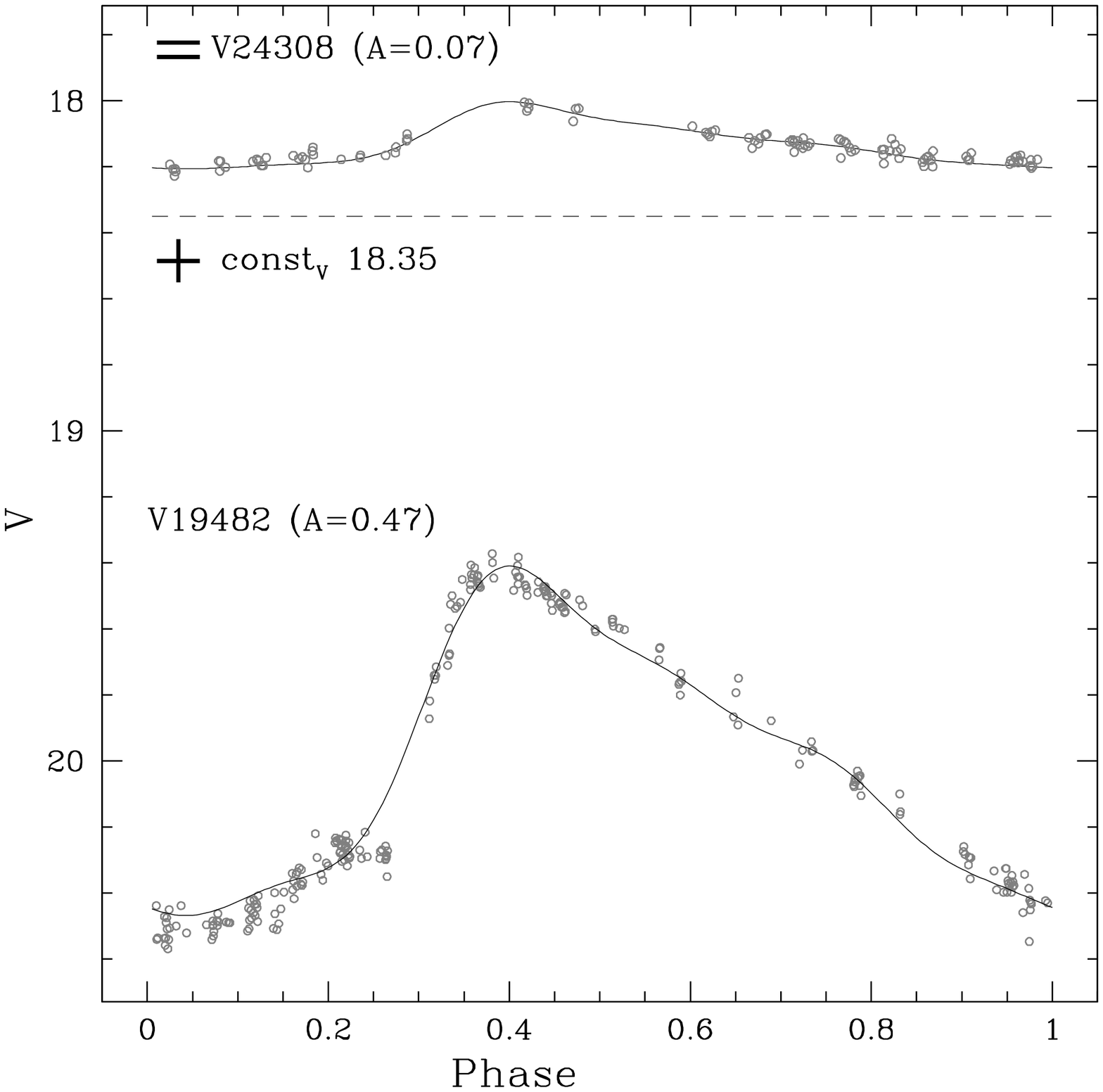}{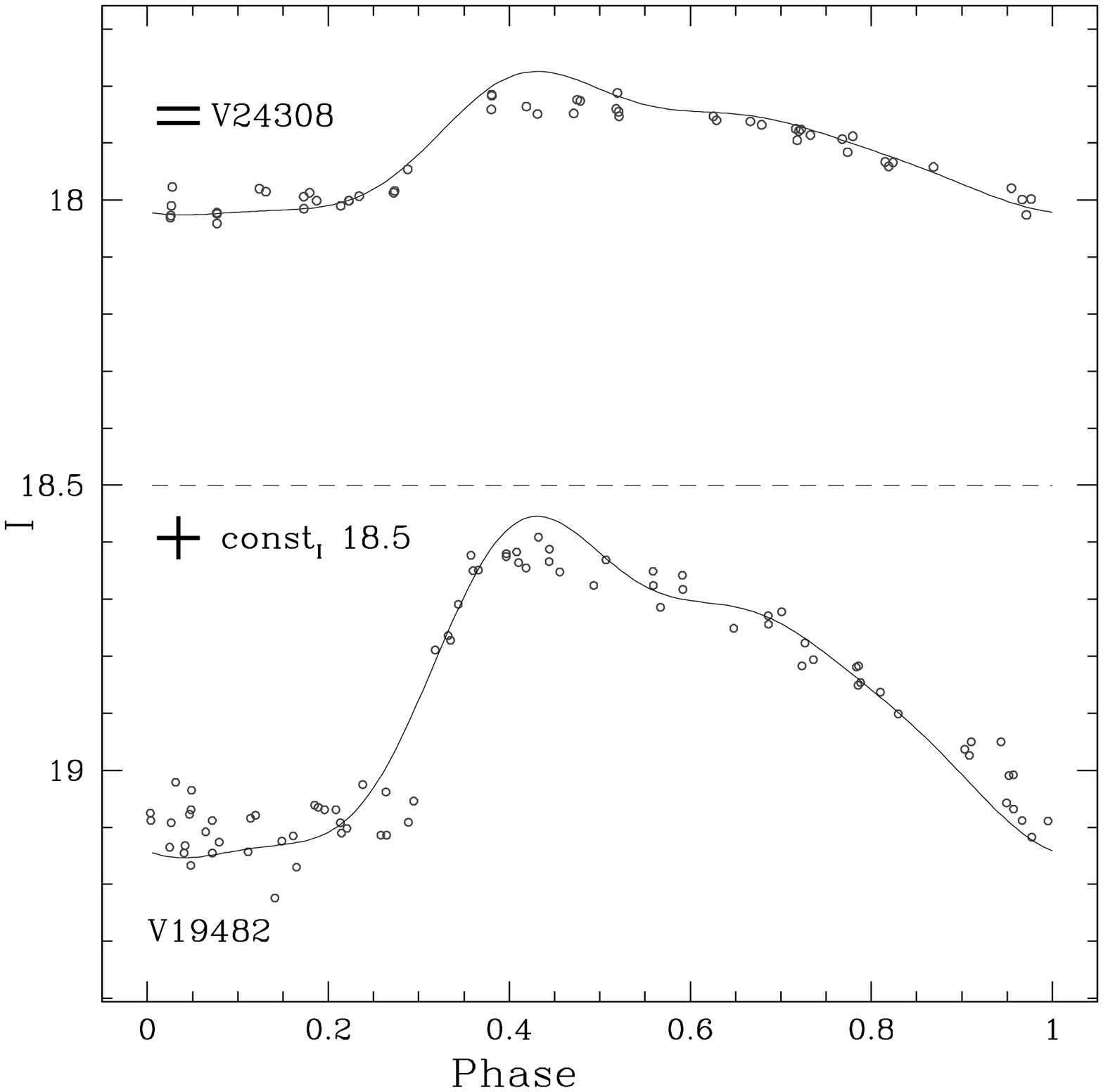}
\caption{A comparison of the $VI$ light curves of the extremely
blended Cepheid D33J013341.3+303212.7 (V24308) and the unblended
Cepheid D33J013324.7+303517.2 (V19482) with a similar period.}
\label{fig:ext}
\end{figure}

\subsection{ The extremely blended Cepheid D33J013341.3+303212.7}

In our ground-based M33 data we have discovered a highly blended
Cepheid, exhibiting a very low amplitude of variability and the color
of an upper main sequence star (D33J013341.3+303212.7, $P=20.15^d$).
We present this Cepheid as more of a curiosity or an instructive
example than a problem in the determination of distances. Such
Cepheids can be very easily rejected on the basis of their low
amplitudes and often discrepant colors.

Without the a priori knowledge of the degree of contamination by
blending, we have estimated the brightness of the companion in $V$ and
$I$ by adding constant flux to the $VI$ light curves of a Cepheid with
a very similar period ($P=20.51^d$, $V=19.96$, $V-I=1.10$), to achieve
the best match with the light curve of the highly blended Cepheid
(Fig. \ref{fig:ext}). From the above analysis we have obtained a $V$
magnitude of 18.35 mag and $V-I$ color of -0.15 mag for the blend.

An investigation of this variable in the HST data yielded the values
of $S_V=3.10$ and $S_B=6.60$. Taking into account the phase at which
the Cepheid was at the time of the HST observation, the blending
companion is brighter by 1.53 mag than the Cepheid in the $V$-band.
In $B$ this difference amounts to about 2 mag. The 1.53 mag difference
in $V$ agrees very well with the 1.61 mag derived from light curve
analysis. 

\section{The M33 Blending Catalog}
We have adopted three criteria that a companion to a Cepheid has to
fulfill to be included into our catalog as a blend. The star has
to:
\begin{enumerate}
\item be located at a distance less than $0\farcs75$ from the Cepheid, 
\item be undetected by DAOPHOT in our ground-based images,
\item contribute at least 6\% of the flux of the Cepheid in the same
filter.
\end{enumerate}
The choice of maximum distance was motivated by the typical full width
at half maximum (FWHM) in our ground-based images ($\sim 1\farcs5$).
We have increased the minimum flux of the blend from 4\% of the flux
of the Cepheid, as adopted in our study of blending in M31 (Paper I)
to 6\%, to obtain a little more conservative estimate.

We have examined the Cepheids on the {\em HST} images to check for
false detections (cosmic rays in case of single images, bad columns,
etc.) or companions missing from the DAOPHOT list. In the former case
the object was removed from the photometry list, in the latter its
coordinates were entered by hand and DAOPHOT was run again on the
corrected list.

To quantify blending we have used the parameter $S_F$, defined in
Paper I as the sum of all flux contributions in filter $F$ normalized
to the flux of the Cepheid:
\begin{equation}
S_F= \sum_{i=1}^{N_F}\frac{f_i}{f_C}
\label{eq:sv}
\end{equation}
where $f_i$ is the flux of the i-th companion, $f_C$ the flux of the
Cepheid on the {\em HST} image and $N_F$ the total number of
companions. 

We have also determined the surface brightness around the Cepheids on
the ground-based $V$-band images taken on the same night as the
photometric standards. The surface brightness was computed as the mode
within a 20 pixel radius. We used a bin width of 1 ADU for the
histogram to compute the value of the mode, smoothed with a
flat-topped rectangular kernel (boxcar) filter 11 units in length.
After correcting for the sky level, the instrumental surface
brightness values were converted to mag/\sq\arcsec using 7-8 fairly
bright isolated reference stars with known standard magnitudes. The
$rms$ scatter around the value of the average ranged from 0.042 to
0.071 mag/\sq\arcsec.

In Table \ref{tab:cep} we present the blending catalog for 102
Cepheids found on the {\em HST} images: the name, the period, the mean
$V$, $I$ and $B$ magnitudes taken from Macri et al. (2001b) and Stanek
et al. (2001), the number of companions $N_F$ and their total flux
contribution $S_F$ in the $V$, $I$ and $B$ bands respectively. The
$V$, as in $N_V$ and $S_V$, refers to filters F555W and F606W, $I$ to
F814W and $B$ to F439W and F450W. For Cepheids identified on more than
one {\em HST} image, the average values of $S_F$ are listed. The last
column gives the $V$ band surface brightness $\Sigma_V$ within a 20
pixel (6\farcs4) radius.

\section{Discussion of the Properties of Blending}

\subsection{The Magnitude of the Effect of Blending }
\label{sub:mag}
\begin{small}
\tablenum{2}
\begin{planotable}{lllllllllllll}
\tablewidth{0pt}
\tablecaption{\sc The M33 Cepheid Blending Statistics}
\tablehead{\colhead{Period} & \colhead{} & \colhead{}& \colhead{$S_V$} & 
\colhead{} & \colhead{} &\colhead{} & \colhead{$S_I$} & \colhead{} &
\colhead{} & \colhead{} & \colhead{$S_B$} & \colhead{} \\ \colhead{range} &
\colhead{} & \colhead{$avg$} & \colhead{$med$} & \colhead{$N$}&
\colhead{} & \colhead{$avg$} & \colhead{$med$} & \colhead{$N$}&
\colhead{} & \colhead{$avg$} & \colhead{$med$} & \colhead{$N$} }
\startdata
all Periods   & & 0.24 & 0.14 & 95 & & 0.30 & 0.21 & 62 & & 0.29 & 0.15 & 57 \\
P $<$ 10 days & & 0.37 & 0.25 & 35 & & 0.43 & 0.29 & 20 & & 0.47 & 0.26 & 18 \\
P $>$ 10 days & & 0.16 & 0.07 & 60 & & 0.23 & 0.14 & 42 & & 0.20 & 0.10 & 39 \\
\enddata
\label{tab:stat}
\end{planotable}
\end{small}

\begin{figure}[t]
\plotfiddle{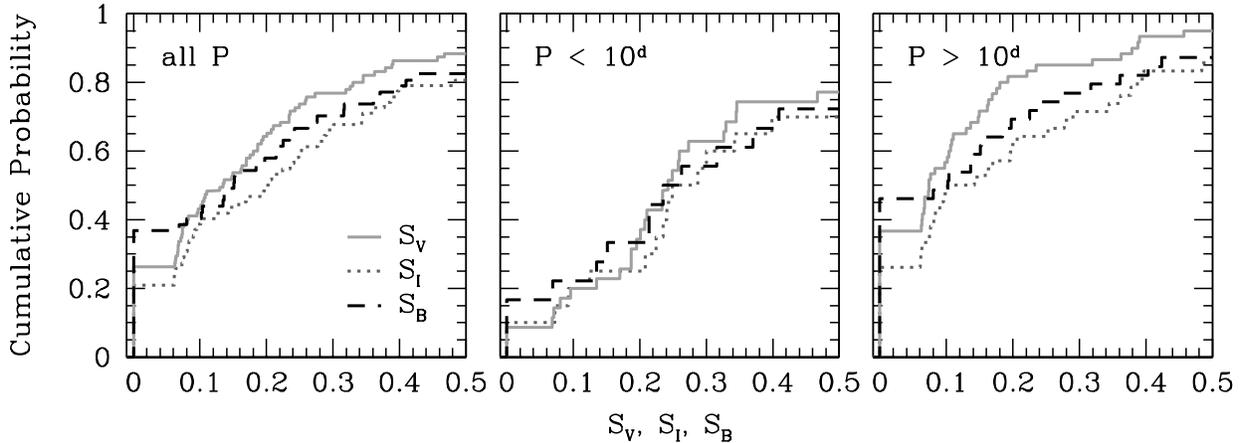}{4.5cm}{0}{85}{85}{-255}{-430}
\caption{The cumulative probability distributions of $S_V$, $S_I$ and
$S_B$ for M33 Cepheids (the continuous, dotted and dashed lines,
respectively). The panels, going from left to right, show the
distribution for the entire sample, for Cepheids with periods below
and above 10 days, respectively.}
\label{fig:ksmag}
\end{figure}

In Table \ref{tab:stat} we present blending statistics for the M33
Cepheids. We list the average and median blending in $BVI$ filters for
the entire sample and for two subsamples, consisting of the Cepheids
with periods below and above 10 days, respectively. The sizes of the
samples ($N$) are also given. For all three samples blending is most
significant in the $I$-band (21\% all P, 29\% $P<10^d$, 14\%
$P>10^d$), intermediate in $B$ (15\%, 26\%, 10\%) and least in $V$
(14\%, 25\%, 7\%). Blending is more substantial for the short period
Cepheids.

Figure \ref{fig:ksmag} shows the cumulative probability distributions
of $S_V$, $S_I$ and $S_B$ for the M33 Cepheids, using solid, dotted
and dashed lines, respectively. The panels, going from left to right,
show these distributions for the entire sample and for Cepheids with
periods below and above 10 days, respectively. The plots are
restricted to Cepheids with blending $S_F$ up to 0.5, to show in more
detail the most interesting range, where cases with blending cannot be
easily distinguished and rejected. About 80-90\% of all Cepheids are
contained within this range.

As expected from the Period-Luminosity relation, the short period
(hence fainter) Cepheids suffer more from blending than do the long
period (hence brighter) ones. Upon examining Fig. \ref{fig:ksmag}, we
again note that the contribution from blends tends to be strongest in
the $I$-band, intermediate in $B$ and weakest in $V$. This trend is
best seen in the long period sample. A likely explanation can be found
by examining the CMDs in Fig. \ref{fig:cmd}. Our limit of $S_F=6\%$
corresponds to a difference of 3.05 magnitudes in brightness between
the Cepheid and its companion. The stars which could contribute enough
flux to become blends will thus most likely be located on the upper
main sequence (MS) or on the red giant branch (RGB). In effect there
should be more companions which are brightest in the $B$-band (upper
MS) or in $I$ (RGB) and relatively few in $V$.

In case of the short period Cepheids, below $S_F=0.25$ ($\sim50\%$ of
the sample) blending is equal across all bands. Two factors could
account for this behavior: poorer statistics due to the smaller size
of the sample, especially in $B$ and $I$, and the increased
contribution of faint intermediate color blends. Above $S_F=0.25$,
$V$-band seems to be least affected by blending.

Another feature seen in this Figure is the relatively large number of
unblended Cepheids in the $B$ band, compared to $I$ and, especially,
$V$. This might be due to the fact that on average the {\it HST}
exposure times in $V$ and $I$ were 2 $-$ 3 times longer than in
$B$. The fainter blends in $B$ could thus elude detection, especially
in case of the short period Cepheids.

\begin{figure}[t]
\plotfiddle{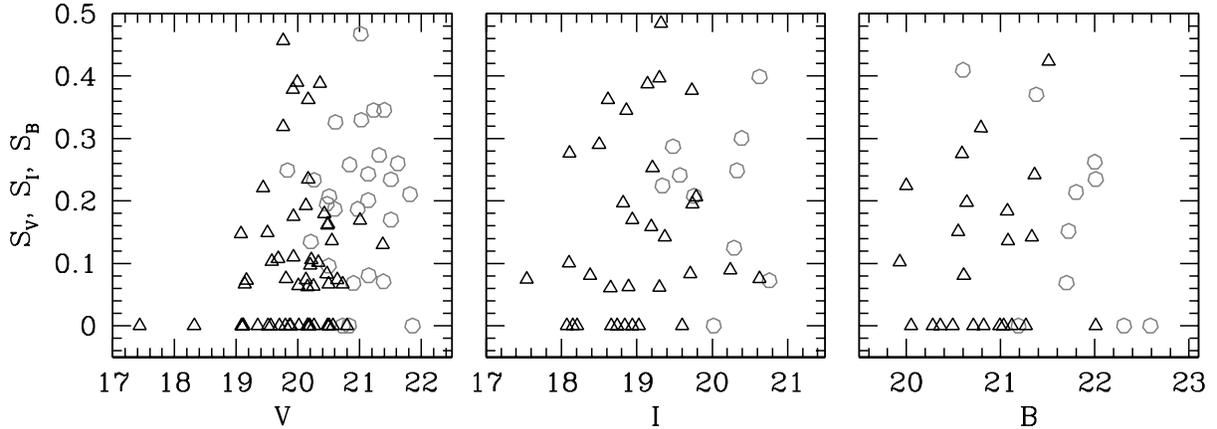}{4.5cm}{0}{85}{85}{-255}{-430}
\caption{The blending parameters $S_V$, $S_I$ and $S_B$ as functions
of the respective ground-based standard mean magnitudes. Short period
Cepheids are indicated with circles, long period ones with triangles.}
\label{fig:vfv}
\end{figure}

In Figure \ref{fig:vfv} we show the blending parameters $S_V$, $S_I$
and $S_B$ as functions of the respective ground-based standard mean
magnitudes. Short period Cepheids are indicated with circles and long
period ones with triangles. No clear correlation of blending with the
magnitude of the Cepheid is seen, although the short period Cepheids
appear on average to be more blended, in accordance with
Fig. \ref{fig:ksmag}.  There is a substantial overlap in brightness for
the short and long period samples.

\begin{figure}[t]
\plotfiddle{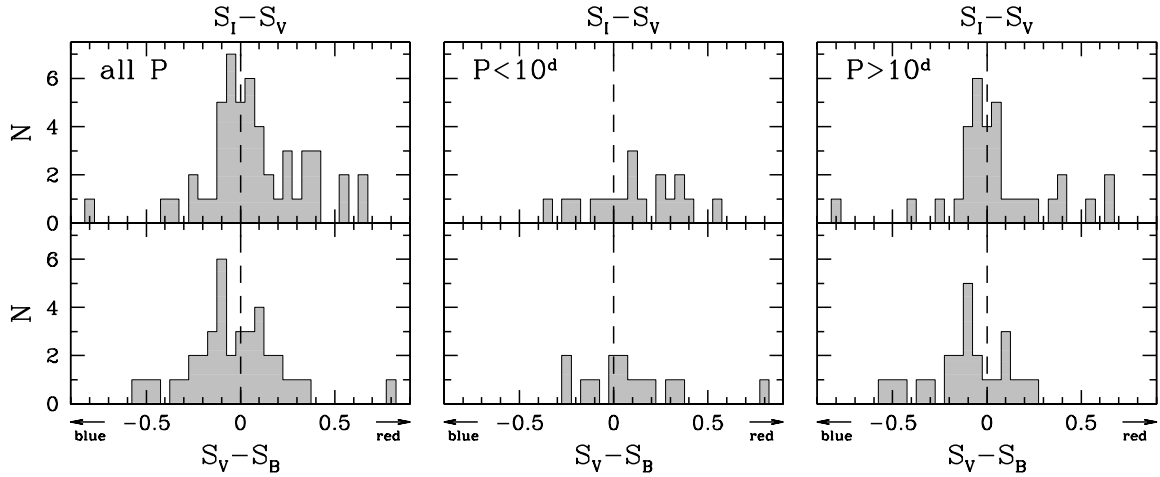}{4.8cm}{0}{85}{85}{-255}{-430}
\caption{ The relative color distributions of the blends in $S_I-S_V$
(top) and $S_V-S_B$ (bottom). The panels, going from left to right,
show the distribution for the entire sample and for Cepheids with
periods below and above 10 days, respectively. }
\label{fig:col}
\end{figure}

\subsection{ The Colors of the Blends}

In Figure \ref{fig:col} we present the relative color distributions of
the blends in $S_I-S_V$ (top) and $S_V-S_B$ (bottom). The panels,
going from left to right, show these distributions for the entire
sample and for Cepheids with periods below and above 10 days,
respectively. Only Cepheids with non-zero blending in at least one of
the bands are plotted. Table \ref{tab:col} lists the median colors and
the number of cases of red and blue blending for each of the three
period ranges.

The color distributions of the blends for the entire Cepheid sample
exhibit maxima approximately at 0. The maxima show double peaks due to
the fact that most blends are either redder or bluer than the
Cepheids.  In the $S_V-S_B$ distribution there are almost as many red
blends as there are blue, while in $S_I-S_V$ there is a noticeable
excess of red blends.

After decomposing the sample into long and short period Cepheids it
becomes apparent that the sample of short period Cepheid blends is
markedly the redder of the two. This is at least in part due to the
correlation between Cepheid luminosity and color -- fainter Cepheids
are bluer than brighter ones. For the long period Cepheids blue blends
dominate in $S_V-S_B$, where there are twice as many blue blends as
red. The $S_I-S_V$ color distribution is almost uniform. The peak is
shifted slightly to the blue, but a red tail is present. The double
structure of the maxima is more prominent than in the entire sample,
especially in $S_V-S_B$.

\begin{small}
\tablenum{3}
\begin{planotable}{lcrcccrcc}
\tablewidth{0pt}
\tablecaption{\sc The M33 Cepheid Blend Color Statistics}
\tablehead{
\colhead{Period} & \colhead{} & \colhead{}& \colhead{$S_I-S_V$} & 
\colhead{} & \colhead{} &\colhead{} & \colhead{$S_V-S_B$} & \colhead{} \\
\colhead{range}      & \colhead{} & \colhead{$med$} & \colhead{$N_{red}$} &
\colhead{$N_{blue}$} & \colhead{} & \colhead{$med$} & \colhead{$N_{red}$} &
\colhead{$N_{blue}$}}
\startdata
all Periods   & & 0.05 & 31 & 20 & & -0.02 & 17 & 21 \\
P $<$ 10 days & & 0.10 & 13 & ~5 & &  0.08 &  9 & ~5 \\
P $>$ 10 days & & 0.02 & 18 & 15 & & -0.09 &  8 & 16 \\
\enddata
\label{tab:col}
\end{planotable}
\end{small}

\subsection{ The Environments of the Cepheids}
\label{sub:sky}
\begin{figure}[t]
\plottwo{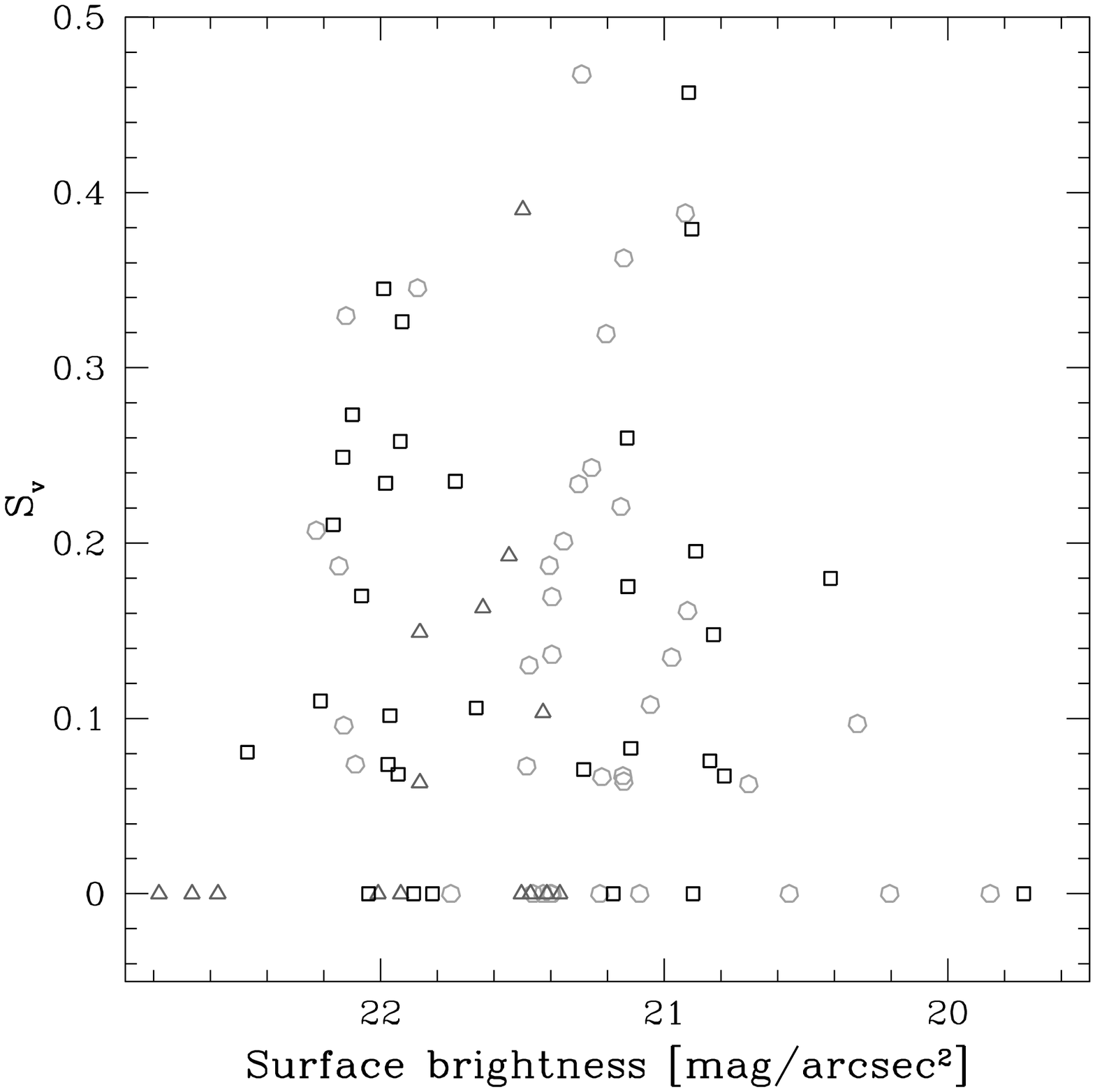}{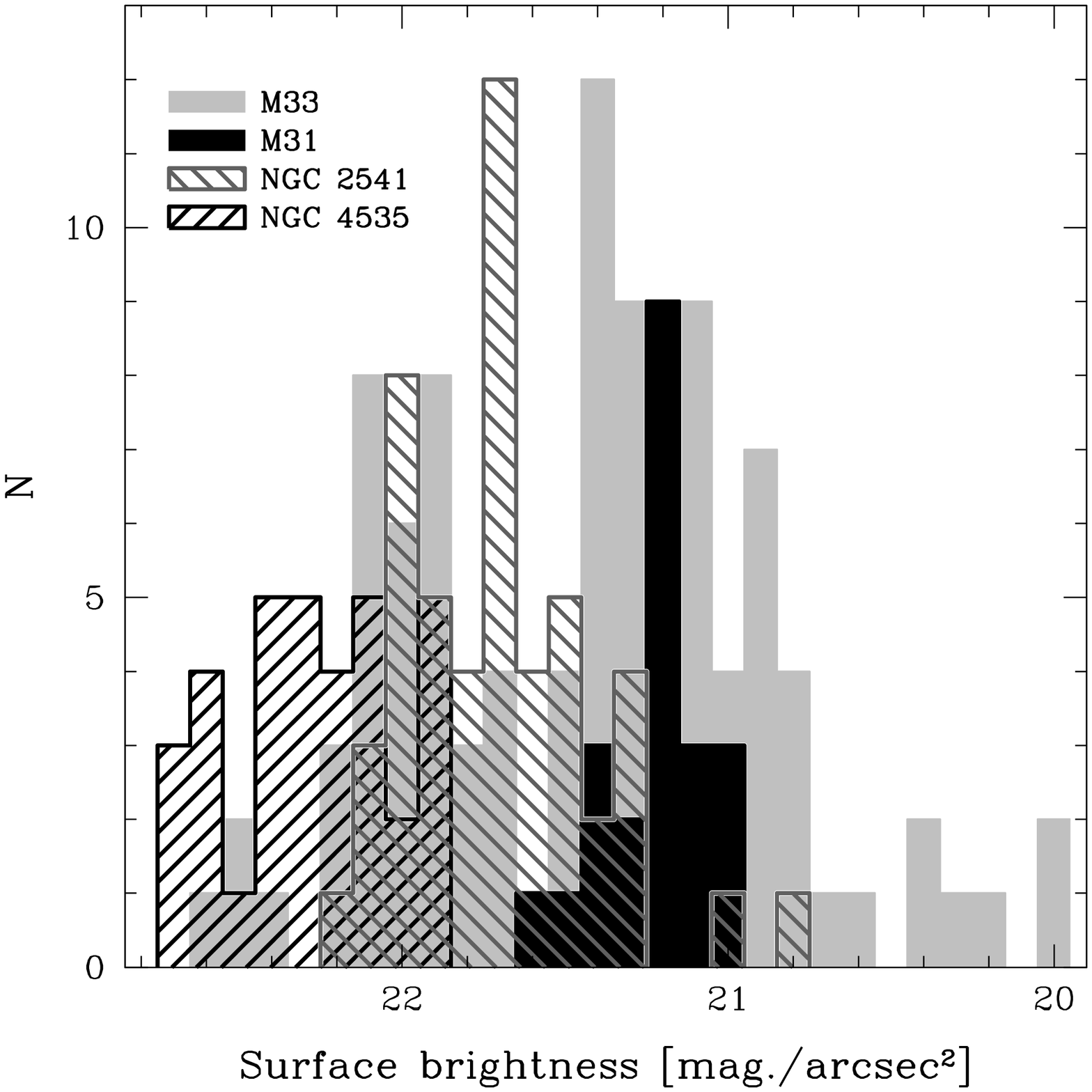}

\caption{Left panel: Blending $S_V$ as a function of the surface
brightness within a radius of $6\farcs4$ around the M33 Cepheids on
the ground-based images. Field A Cepheids are denoted by squares,
field B by circles and field C by triangles.  Cepheid blending does
not seem to correlate with surface brightness. Right panel: The
surface brightness around Cepheids in M33, M31, NGC 4535 and NGC 2541. }
\label{fig:sky_sv}
\label{fig:sky_sv_cmp}
\end{figure}

There have been suggestions in the literature (e.g. Ferrarese et
al. 2000) that blending should increase with the measured values of
surface brightness. We investigate empirically this effect using our
M33 data. The left panel of Figure \ref{fig:sky_sv} shows blending
$S_V$ as a function of the $V$-band surface brightness within a radius
of $6\farcs4$ around the M33 Cepheids on the ground-based
images. Field A Cepheids are denoted by squares, field B by circles
and field C by triangles.  We find no such correlation between
blending and the underlying surface brightness -- Cepheids in
environments of different surface brightness show roughly the same
frequency and amount of blending.

To put these values into perspective, we have computed the surface
brightness for two galaxies which straddle their range, NGC 4535 and
NGC 2541. The {\em HST} data for these galaxies, observed as part of
the HST Key Project on the Extragalactic Distance Scale (Macri et
al. 1999, Ferrarese et al. 1998), were obtained from the Hubble Data
Archive. The implemented procedure was described in Paper I. In the
right panel of Figure \ref{fig:sky_sv_cmp} we plot the surface
brightness distributions for Cepheids in M33, M31 (Paper I), NGC 4535
and NGC 2541. As seen in the diagram, M33 spans a wide range in
surface brightness -- three magnitudes, from 19.7 to 22.8
mag/\sq\arcsec. The distribution shows two clear maxima: the brighter
one corresponds to Cepheids located within the spiral arms and the
fainter one to those in regions in between the arms. There is also a
tail of Cepheids from very bright regions, in the proximity of the M33
nucleus. M31 overlaps mostly with the brighter maximum of M33. The NGC
4535 distribution peaks slightly above the fainter maximum of M33. NGC
2541 also overlaps the fainter maximum in M33, though about half of
its Cepheids are located in regions of lower surface brightness, where
our M33 coverage is much more sparse (4 Cepheids). We conclude that
our Cepheids in M33 reside in environments of surface brightness
typical of spiral galaxies.

\section{ Crowding vs. Blending - Artificial Star Tests}
In our previous paper on blending of Cepheids in M31 (Paper I) we have
made a distinction between the phenomena of crowding and blending. We
defined {\em blending} as the close projected association of a Cepheid
with one or more intrinsically luminous stars, mostly other young
stars which are physically associated - from actual binary and
multiple systems to companions which are not gravitationally bound to
the Cepheid. In the case of our investigation we are limited by
resolution to wide systems. {\em Crowding} or {\em confusion
noise} occurs in stellar fields with a crowded and complex background
due to the random superposition of stars of different luminosity. 

In such dense fields as are observed in galaxies, the blending estimate
for a Cepheid will also include crowding. Based on our M33 data we
will try to address the issue of the importance of blending relative
to crowding. If we assume that a Cepheid is associated with other
luminous stars located in its proximity, then moving it to a randomly
chosen position on the image will break that association. In the
former case, the Cepheid will be subject to blending; in the latter,
to crowding.

To estimate the influence of crowding we have generated a catalog of
randomly distributed stars with the same magnitudes as the actual
Cepheids, following the same technique as for the Cepheid blending
catalog. For each Cepheid observed on a WFPC2 image we generated a
list of 100 random positions and determined the contribution from
companions at that location. We also determined the surface brightness
values on the ground-based images around these positions. Our
artificial star catalog, thereafter referred to as the {\em
artificial} catalog, contains data for $\sim9200$ realizations in $V$,
$\sim5300$ in $I$ and $\sim 3000$ in $B$.

The construction of the blending catalog described in Section 4
(thereafter referred to as {\em corrected}) involved the visual
examination of all putative companions, to correct for missing and/or
false detections. The large number of realizations in the artificial
crowding catalog precluded us from making a similar inspection.
Therefore, the original unaltered catalog of Cepheid blends
(thereafter {\em uncorrected}) seems to be the more natural choice for
a comparison between blending and crowding. In the discussion that
follows, we have used the uncorrected blend catalog; K-S tests show
that at confidence levels of 93\%-100\% the distributions for the
corrected and uncorrected catalogs are consistent with a single
distribution (Tab. \ref{tab:ks_test}).

\begin{figure}[t]
\plotfiddle{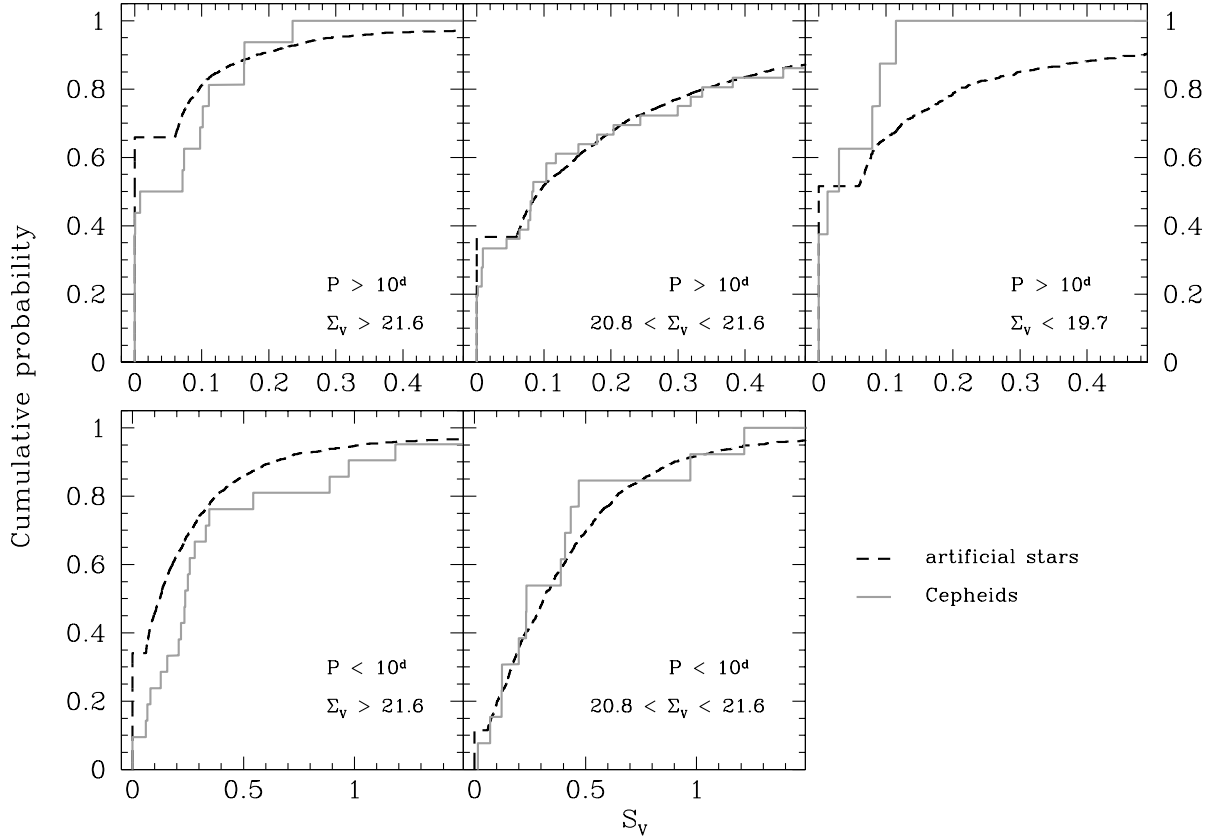}{10cm}{-90}{60}{60}{-250}{340}
\caption{The cumulative probability distributions of the blending
parameter $S_V$ for the artificial Cepheids (dashed line) and the
uncorrected Cepheid catalog (solid).}
\label{fig:ks_test}
\end{figure}

\subsection{The Magnitude of the Effects of Blending and Crowding}
We have divided the sample of M33 Cepheids into two bins in period at
$P=10^d$ and three bins in surface brightness, separated at
$\Sigma_V=21.6$ and $\Sigma_V=20.8$ mag/\sq\arcsec. The division into
different surface brightness regions was motivated by the clear
separation of Cepheids located inside and outside the spiral arms and
the existence of a high surface brightness tail associated with the
nucleus, as seen in the right panel of Fig. \ref{fig:sky_sv_cmp}. The
magnitude of the effect of crowding will be different in each of the
three ranges of surface brightness, therefore we have decided to
analyze them separately.

Figure \ref{fig:ks_test} shows the cumulative probability
distributions for $S_V$ drawn from the artificial crowding catalog
(dashed line) and the uncorrected Cepheid blending catalog (solid
line). We show the distributions for Cepheids with $P>10^d$ in the
upper panels and with $P<10^d$ in the lower ones. Note that the scales
on the $S_V$ axes are different for the two period ranges. The left,
middle and right panels show the lowest, medium and highest surface
brightness regions, respectively.  The lower right panel is not drawn,
as there was only one short period Cepheid in the highest surface
brightness range.

Let us first consider the Cepheids located in regions of lowest
surface brightness (leftmost panels of Fig. \ref{fig:ks_test}). A
visual inspection reveals that blending is stronger than crowding for
these Cepheids. The K-S test confirms the impression that these data
sets are different, giving a 1\% and 3\% probability that they are
drawn from the same distribution, for the short and long period
Cepheids, respectively. The difference between the magnitude of those
two effects for the long period Cepheids is not as large as for the
short period ones. This is expected, as a short period (fainter)
Cepheid will yield a higher value of $S_V$ than a long period
(brighter) Cepheid when blended with a star of the same magnitude.

We now move on to Cepheids located in intermediate surface brightness
regions (middle panels of Fig. \ref{fig:ks_test}). At first glance, it
appears that the magnitude of the effects of blending and crowding is
comparable. For the short period Cepheids the K-S test shows that
these data sets are consistent with a single distribution at a
confidence level of 93\%. In case of the long period sample, the K-S
test gives a probability of only 10\%. Nevertheless, even if the
distributions differ, the two effects are still similar in magnitude.

Finally, we examine the case of Cepheids located in the highest
surface brightness regions (upper right panel of Fig. \ref{fig:ks_test}). 
We note that blending is weaker than crowding for these Cepheids. This
effect is most likely due to selection effects. In regions of such
high surface brightness we are biased towards finding luminous
Cepheids with weaker than average blending. The K-S test results are
not conclusive in determining whether or not these two data sets are
drawn from different distributions. Also, crowding for these Cepheids
is weaker than for those in the intermediate surface brightness
regions, as luminous Cepheids are less succeptible to crowding.

The above analysis indicates that the importance of blending relative
to crowding very likely increases with decreasing surface brightness.
This is not unexpected, as young stars are known to cluster (Harris \&
Zaritsky 1999). Increasing the level of crowding will tend to obscure
this effect.

\begin{small}
\tablenum{4}
\begin{planotable}{lcrrr}
\tablewidth{0pc}
\tablecaption{\sc The K-S Test Results }
\tablehead{
\colhead{Period} & \colhead{$\Sigma_V$} & 
\colhead{$P_{KS}^{C,U}$} & \colhead{$P_{KS}^{U,A}$} & \colhead{$N_C$}}
\startdata
$P<10^d$ & $> 21.6$       & 100.0 &  0.9 & 21 \\
         & $20.8\div21.6$ &  99.5 & 93.0 & 13 \\
$P>10^d$ & $> 21.6$       & 100.0 &  3.4 & 16 \\
         & $20.8\div21.6$ &  97.1 &  9.9 & 36 \\
         & $< 20.8$       &  92.9 & 31.0 &  8 \\
\enddata
\tablecomments{C - the corrected Cepheid blending catalog
(Tab. \ref{tab:cep}); U - the original uncorrected Cepheid blending
catalog; A - the artificial Cepheid crowding catalog.}
\label{tab:ks_test}
\end{planotable}
\end{small}

\subsection{ The Colors of Blending and Crowding Companions}

\begin{figure}[t]
\plotfiddle{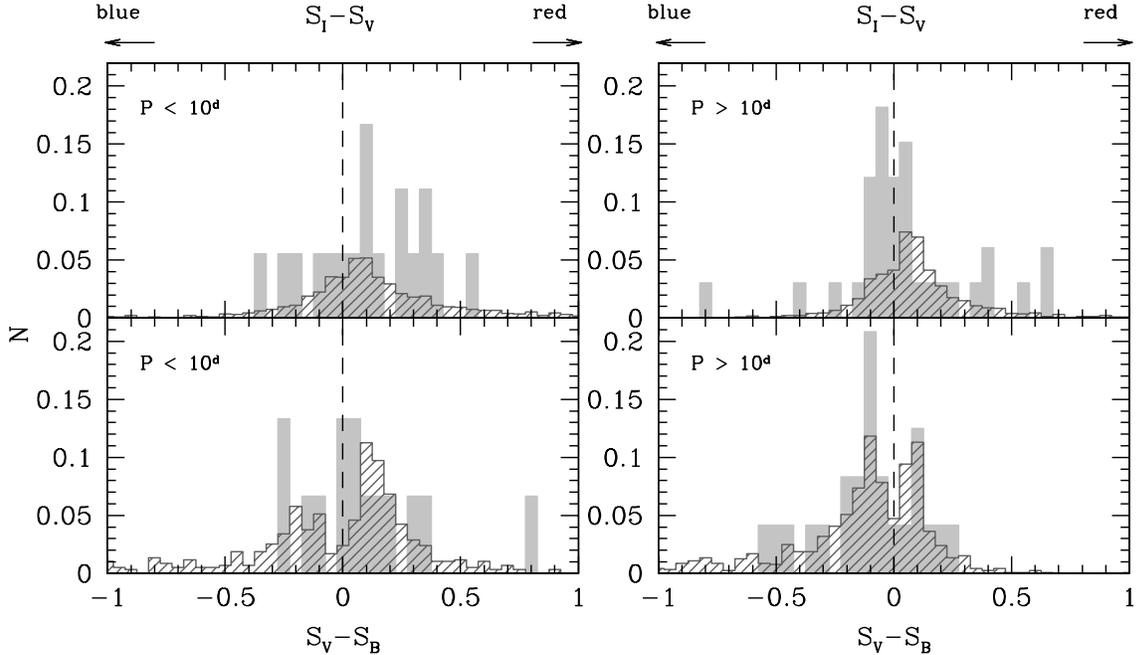}{7.5cm}{0}{80}{80}{-250}{-140}
\caption{The relative color distributions of the blending and crowding
companions in $S_I-S_V$ (top) and $S_V-S_B$ (bottom), drawn from the
artificial crowding catalog (hatched histogram) and the corrected
Cepheid blending catalog (filled histogram). The panels, going from
left to right, show the distribution for Cepheids with periods below
and above 10 days, respectively. The distributions are normalized to
unity.}
\label{fig:ast_col}
\end{figure}

Figure \ref{fig:ast_col} presents the relative color distributions of
the blending and crowding companions in $S_I-S_V$ (top) and $S_V-S_B$
(bottom), drawn from the artificial star crowding catalog (hatched
histogram) and the corrected Cepheid blending catalog (filled
histogram). We only plot the distribution from the corrected catalog,
since it is closer to the true color distribution than the one drawn
from the uncorrected catalog. The panels, going from left to right,
show the distribution for Cepheids with periods below and above 10
days, respectively. Only Cepheids with non-zero blending in at least
one of the bands are plotted. The distributions are normalized to
unity. The number of Cepheids from the corrected and uncorrected
catalogs may be different, as some Cepheids will pass the 6\%
threshold in one catalog and not in the other.

Let us first examine the colors of the crowding companions. The
$S_I-S_V$ color distributions are shifted towards the red, though the
one for long period Cepheids is asymmetric, with a small bump
on the blue side. In $S_V-S_B$ the distributions show two peaks. For
short period Cepheids the red peak is more pronounced, while for
long period Cepheids they are comparable and there is a tail extending
bluewards.  This is consistent with the interpretation presented in
\S\ref{sub:mag}. The stars which could contribute enough flux
to cross the 6\% threshold will most likely be blue stars on the upper
MS or red stars on the RGB.

As far as can be inferred from the modest sample of short period
Cepheids, the blends to the Cepheids show qualitatively similar color
distributions as the companions due to crowding. The color
distributions of long period Cepheid blends show subtle differences
from the crowding color distributions. The $S_I-S_V$ color
distribution of blends with long period Cepheids peaks to the blue of
the distribution of crowding companions, while in $S_V-S_B$ the blue
peak is more distinct than the red one, while for the artificial stars
the two peaks were of comparable magnitude.

The above comparison indicates that the distribution of the colors of
blends to the short period Cepheids and of crowding stars are most
likely similar, while the blends to the long period Cepheids tend to
be bluer than the crowding stars. If so, this would lend support to
the notion that more luminous (hence younger) Cepheids should be
located close to their place of formation, near other young blue
stars.

\section{ Indications for Remote Galaxies}

As the number of extragalactic Cepheids discovered with {\em HST}
continues to increase, it becomes crucial to obtain constrains on the
bias in their photometry introduced by the effect of blending.  Using
the HST M33 data as the template we will try to obtain an estimate of
the effect that blending would have on this galaxy if it were observed
at their distances. The actual magnitude of the effect of blending
will depend on many factors, such as the morphology of the host
galaxy, the surface brightness of the regions in which the Cepheids
are located, the brightness distribution of the Cepheids, the actual
sample of Cepheids, the methods of variable extraction and
classification, etc. Nonetheless, M33 can yield a useful order of
magnitude constraint, as it is a spiral galaxy of surface brightness
typical for those where Cepheids are sought and is observed at a
moderate inclination ($56^\circ$, Schmidt, Priebe \& Boller 1993).

\begin{figure}[t]
\plotfiddle{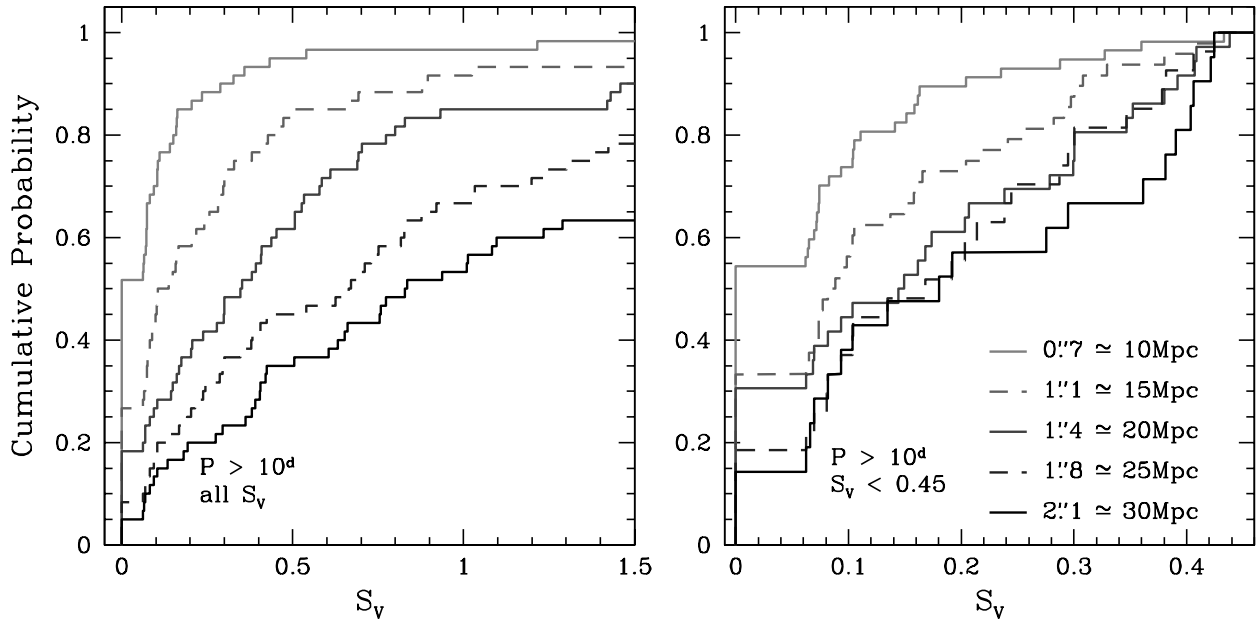}{7cm}{0}{90}{90}{-270}{-170}
\caption{The cumulative probability distributions of blending
parameter $S_V$ for the M33 Cepheids within radii of $0\farcs7$,
$1\farcs1$, $1\farcs4$, $1\farcs8$, $2\farcs1$, corresponding
to $0\farcs12$ resolution at distances of 10, 15, 20, 25 and 30 Mpc,
assuming a distance of 850 kpc to M33.}
\label{fig:rem}
\end{figure}

\begin{figure}[t]
\plotfiddle{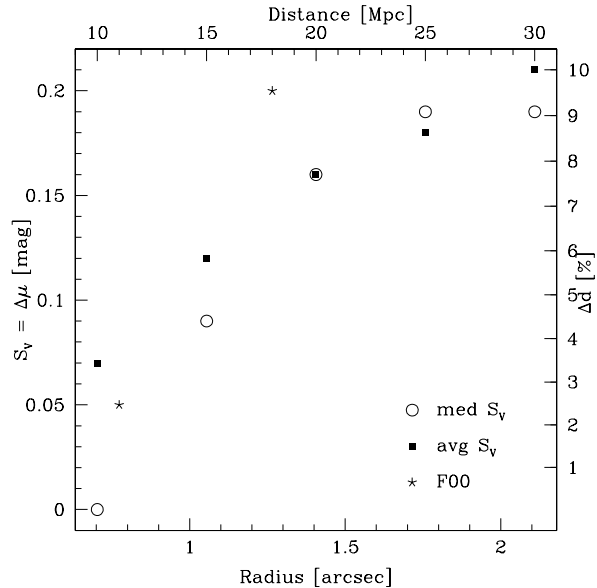}{6.5cm}{0}{40}{40}{-130}{-70}
\caption{Blending/distance bias as a function of the summing
radius/distance. The average and median $S_V$ are indicated with
open and filled symbols, respectively. The Ferrarese et al. (2000)
results are indicated with asterisks.}
\label{fig:bvd}
\end{figure}

The average FWHM on the DIRECT project ground-based images of M33 is
about $1\farcs5$, or $\sim 6$ pc. A similar spatial resolution would
be achieved with the WF chips of the {\em HST}-WFPC2 camera for a
galaxy at a distance of $11$ Mpc. By increasing the radius around the
Cepheid for summing the contributions of the blends we can simulate
the deterioration of resolution due to the increasing distance to the
galaxy. For our purpose we have chosen the radii of $0\farcs7$,
$1\farcs1$, $1\farcs4$, $1\farcs8$, $2\farcs1$, corresponding to the
$0\farcs12$ FWHM on the WF chips at distances of 10, 15, 20, 25 and
30 Mpc, assuming a distance of 850 kpc to M33. We have restricted
ourselves to the long period ($P>10^d$) sample, as long period
Cepheids are preferred for determining distances.

The left panel of Figure \ref{fig:rem} shows the cumulative
probability distribution of $S_V$ for different cutoff radii (proxy
for distance). As expected, blending increases with distance and
becomes severe at radii corresponding to M33 seen at distances of 20,
25 and 30 Mpc, where the median 0 of $S_V$ are 37\%, 68\% and
89\%. The highly blended Cepheids will of course be recognized as such
based on the diminished amplitude of variability and, possibly,
different colors.

In a study of crowding Ferrarese et al. (2000) find that it is
virtually impossible to recognize a Cepheid with $S_V\leq 30\%$ as
affected from the change in the amplitude of the light curve, and it
only becomes possible for most Cepheids when $S_V$ exceeds 60\%. To
make our estimate of blending more realistic we have redrawn these
distributions in the right panel of Fig. \ref{fig:rem} making the
assumption that all Cepheids with $S_V>45\%$ will be recognized as
blended and rejected from the sample. Here blending also increases
with distance for the most part, though the distributions are seen to
cross each other on several occasions. Most striking is the case of
the 25 Mpc and 30 Mpc distributions, where strong blending
($S_V>20\%$) is more severe at 25 Mpc than at 30 Mpc. This is due to
the fact that as the overall value of blending increases, the slope of
the $S_V$ distribution flattens out and does not change considerably
in the relevant range of $S_V$.

The bias in distance due to blending is illustrated in
Fig. \ref{fig:bvd} as a function of distance.  At distances of 10-20 Mpc
the bias increases from $0\%-3\%$ to $8\%$, depending on the statistic
used and levels off at $9\%-10\%$ at 25-30 Mpc. The bias introduced by
blending into the distance modulus expressed in magnitudes is to a
very good approximation equal to $S_V$ for $S_V \leq 0.2$. 

\section{Other investigations of blending}
\label{sect:oth}

Since the appearance of our first paper based on M31 Cepheids (Paper
I) the effect of blending on the Cepheid distance scale has been
subject to further study. In this section we will present a short
discussion of the results obtained by other groups and try to relate
our results to theirs. In several papers our results and those of
Stanek \& Udalski (1999) discussed below, have been questioned; we
will also address this issue.

Stanek \& Udalski (1999) estimated the effect of blending for remote
galaxies using the OGLE LMC Cepheid and star catalogs (Udalski et
al. 1999, 2000) as the unblended template. Their approach involved summing
the contribution of blends within a range of radii around the
Cepheids, corresponding to the resolution of the WF camera at
different distances. The distance moduli were derived from fitting the
P-L relations to all Cepheids whose $I$-band amplitudes exceeded 0.4
mag after accounting for blending. Their results indicated that
blending could introduce a substantial bias into Cepheid derived
distances: $<$0.1 mag at distances $<$15 Mpc and 0.2-0.3 mag for more
remote galaxies. 

An analysis of the effect of confusion noise on Cepheid distances
based on artificial star tests was presented by Saha, Labhardt \&
Prosser (2000). That work dealt with the contribution from the
underlying confusion pattern resulting from the unresolved stellar
background. Their simulations showed that this effect is within 0.1
mag for NGC 4639 at a distance of 25 Mpc. Our estimates include both
confusion noise and stars that would be resolved if they were observed
at a larger separation from the Cepheid. Therefore it is not
surprising that the Saha et al. result is somewhat lower than our
estimate of a 0.2 mag bias at a distance of 25 Mpc. 

A further study of the effects of crowding and confusion noise,
addressing specifically the case of the HST Key Project, was performed
by Ferrarese et al. (2000) using artificial star tests. They estimated
that the effect of crowding could bias the photometry of the Cepheids
to be too bright by 0.05 mag for NGC 2541 at 11 Mpc and 0.2 mag for
NGC 1365 at 18 Mpc.

The Ferrarese et al. simulations account only for the effect of
crowding, and not for blending, which appears to be more significant
than crowding in low surface brightness regions
(Fig. \ref{fig:ks_test}). This dependence will most likely result from
the clustering of Cepheids with other young stars and will become
more prominent at lower levels of crowding. Such clustering has been
observed for blue MS stars in the LMC (Harris \& Zaritsky 1999).

\begin{figure}[t]
\plotfiddle{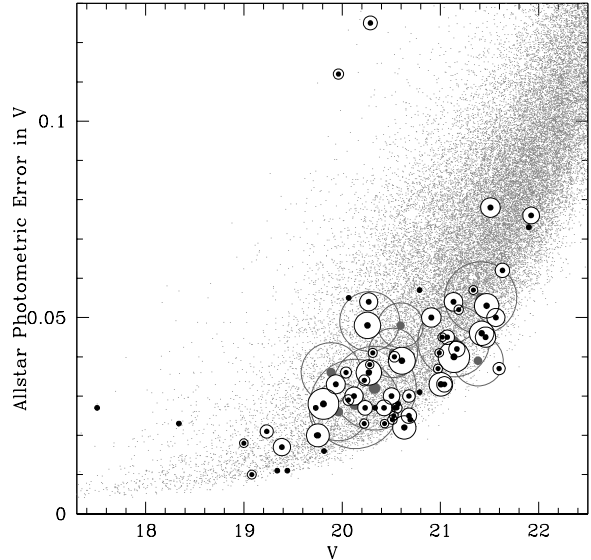}{6cm}{0}{40}{40}{-140}{-70}

\caption{The Allstar photometric errors for the M33 field A and
B stars as a function of the standard $V$-band magnitude. The
Cepheids are indicated by solid dots surrounded by circles
proportional in size to $S_V$ (no outside circle means $S_V=0$). The
white-filled symbols indicate Cepheids with $S_V<0.5$}
\label{fig:psf_err}
\end{figure}

Ferrarese et al. assume that most of the affected Cepheids can be
rejected using multiple epoch data on the basis of several criteria,
thus diminishing the bias in distance to less than 0.02 mag. One of
these criteria is based on the assumption that affected stars should
have larger photometric errors. We have investigated the existence of
a correlation between Cepheid blending and the reported
DAOPHOT/ALLSTAR photometric errors. In Figure \ref{fig:psf_err} we
plot the photometric errors of stars in M33 fields A and B as a
function of magnitude, from Macri et al. (2001c). Cepheids are
indicated by solid dots surrounded by circles of size proportional to
their value of $S_V$. The white-filled symbols indicate Cepheids with
$S_V<0.5$. This figure clearly illustrates that there is no
correlation between photometric errors and $S_V$ -- any kind of cut
based on photometric errors will not discriminate between affected and
unaffected Cepheids. 

Another selection criterion examined by Ferrarese et al. (2000) is
that a Cepheid to be used in the distance determination should fall
within the magnitude range applicable to its period range. In practice
this amounts to the rejection of outliers deviating from the mean by
more than 3 times the $\sigma$ of the best fit (Ferrarese et
al. 1998). In case of NGC 2541 at 12 Mpc the 3 $\sigma$ deviation
corresponds to $\pm0.81$ mag in $V$ and $\pm0.54$ mag in $I$. Even
seriously affected Cepheids can pass this criterion. Ferrarese et
al. (2000) confirm that it is virtually impossible to discriminate
against Cepheids for which the contamination amounts to 30\% or less
of the Cepheid mean flux. These are the most common types of blends,
and thus of greatest concern (only 15\% of our $P>10^d$ Cepheids have
$S_V > 30\%$).  Furthermore, they find that only when the contamination
is increased above 60\% are they able to recognize most Cepheids as
affected.

We do not observe the correlation between blending and surface
brightness, predicted by Ferrarese et al. (Fig. 6 in Paper I;
Fig. \ref{fig:sky_sv}). Contrary to their unsupported claim the M31
Cepheids used in our analysis are not located in regions where the
stellar background is significantly brighter than for the HST KP
galaxies (Fig. 7 in Paper I; Fig. \ref{fig:sky_sv_cmp}). Lastly, based
on the residuals from the $H$-band Tully-Fisher relation Ferrarese et
al. concluded that the blending hypothesis could be ruled out at the
1.85 $\sigma$ level. As Gibson et al. (2000) pointed out, this result
is erroneous, as they had neglected the uncertainties associated with
each of the residuals.

Three empirical tests of blending were presented by Gibson, Maloney \&
Sakai (2000). The predicted influence of blending on Cepheid derived
distances was quantified by fitting a functional form to the results
for LMC Cepheids presented by Stanek \& Udalski (1999), which, as
indicated therein, seems to have on average higher surface brightness
than a typical HST Key Project galaxy. The authors find that the
observed distribution of peak luminosities for five type Ia SNe
excludes the blending hypothesis defined as above at a $\sim$1
$\sigma$ level. The second test, $I$ and $H$-band Tully-Fisher
residuals, is not able to discriminate between the no blending and
strong blending hypothesis: both models differ from the best fit
lines, at most, at the 1 $\sigma$ level. The third test involves the
comparison of distance moduli derived separately from the PC chip and
the WF chips. As the PC chip has twice the resolution of the WF chips,
there should be a systematic difference between these distance
moduli. For five galaxies at a distance of $\sim31$ mag, the observed
differences have an rms scatter of 0.08 mag around the mean value of
0.03 mag, while the predicted offset is 0.11 mag. We would advise
caution when drawing conclusions from this comparison. There are many
other factors that the authors have neglected which will invalidate
the results of such a comparison, like poor statistics on the PC chips
(9, 6, 6, 6 and 4 Cepheids for the five galaxies), different
orientations of the camera with respect to the galaxy, selection
effects (see upper right panel of Fig.  \ref{fig:ks_test}), finite
intrinsic width of the P-L relation. Gibson et al. results are
inconclusive and only indicate at a 1 $\sigma$ level that blending for
Cepheids located in the LMC bar is not representative for the distant
galaxies observed by the HST Key Project.

The final HST Key Project paper (Freedman et al. 2001) includes into
the systematic error budget an uncertainty of $^{+5}_{-0}\%$ due to
crowding and blending. The authors conclude that to assess
quantatively the impact of unresolved blending effects would require  
simulations based on the distribution of Cepheids in a galaxy field   
unaffected by blending, such as the study applied to the M101 Cepheids
observed with NICMOS (Macri et al. 2001a).

The choice of M101 as the target to study the effects of blending was
motivated by the large differences in the $H$ and $J$-band
Cepheid-based distance moduli in the inner and outer M101 fields:
$\Delta \mu_H = 0.46\pm0.12$ mag and $\Delta \mu_J = 0.37\pm0.12$ mag.
The approach adopted by Macri et al. (2001a) involved scaling the
stellar separations in the well resolved fields in M31 and M81 to the
distance of M101 and their magnitudes to the exposure time of the M101
images. Artificial fields were generated from these star lists and
Cepheid photometry extracted following the same procedure as for the
actual data. The resulting distance moduli obtained from the Cepheid
PL fits were found to be smaller than the input ones by 0-0.2 mag,
depending on the applied period cutoff. Based on these results the
authors conclude that a substantial fraction of the difference in the
distance moduli could be due to blending. In addition, the fact that
the inner M101 field Cepheids exhibit the same correlation between
$\langle E(V-I)\rangle$ and $\langle E(V-H)\rangle$ as those in other
galaxies suggests that on average, the contamination due to blending
has not introduced a significant change in the color of the Cepheids.

The issue of blending has also been studied for stars in general, to
estimate its influence on the shape of the derived luminosity
functions. At high enough levels of stellar density an individual
resolution element may contain random clumps of several stars, and a
photometric reduction package will not be able to recognize the
multiple nature of the object. Renzini (1998) explored this issue on
theoretical grounds and found that the number of blends is
proportional to the square of both the surface brightness and the
actual resolution. He concluded that meaningful photometry can be
obtained for stars much brighter than the total luminosity observed in
one resolution element. This effect was also investigated empirically
by DePoy et al. (1993), who compared the luminosity functions derived
from images binned by varying factors and by Stephens et al. (2001),
using simulations of entire star clusters, as observed in M31 with
NICMOS. Both groups find that due to extreme crowding, the measured
luminosity function may appear to extend to brighter magnitudes than
the true luminosity function. They also point out that typical
artificial star experiments will not reproduce the effect of crowding
discussed above.

\section{Conclusions}
We have determined the effect of blending in $BVI$ for a sample of 102
Cepheids in the M33 galaxy by comparing our ground-based images
collected with a 1.2 m telescope to archival {\it HST} data. We found
the average (median) flux contribution from luminous companions not
resolved on the ground-based images in the $V$, $I$, $B$ bands ($S_V$,
$S_I$, $S_B$) to be about 24\% (14\%), 30\% (21\%), 29\% (15\%) of the
flux of the Cepheid. For 64 Cepheids with periods in excess of 10 days
the average (median) values of $S_V$, $S_I$, $S_B$ are 16\% (7\%),
23\% (14\%), 20\% (10\%). Blending is least significant in $V$,
intermediate in $B$ and strongest in $I$. These results show that
depending on the sample of Cepheids chosen, our ground-based $V$-band
photometry for the M33 Cepheids could be systematically underestimated
by 8\%-11\% (3\%-6\%). 

We have studied crowding and blending as separate phenomena using
artificial star tests. Our results indicate that the effect of
blending could be more significant than crowding in regions of lower
surface brightness, below 21.6 mag/\sq\arcsec. The two effects are
comparable in magnitude in intermediate surface brightness regions
($20.8\leq \Sigma_V \leq 21.6$ mag/\sq\arcsec).

We have also investigated the colors of the blends and of the crowding
companions. The blends to short period Cepheids ($P<10^d$) are on
average redder than the blends to long period Cepheids ($P>10^d$),
in accordance with the Cepheid period$-$luminosity$-$color relation.
In most cases, blending does not significantly influence the
color of the Cepheid. As in the case of the blends, the crowding
companions to short period Cepheids are on average redder than the
companions to long period Cepheids. We compare the blending and crowding
color distributions and find indications that for long period Cepheids
the blends could on average be bluer than the companions introduced by
crowding.

We have estimated the effect of blending at resolutions corresponding
to a range of distances from 10 to 30 Mpc, using the {\it HST} M33
data as an unblended template stellar population. Assuming that all
Cepheids with $S_V>45\%$ will be recognized as blended and rejected
from the sample, we find that the distance underestimate increases
from $0\%-3\%$ to 8\% at 10-15 Mpc and levels off at 9\%-10\% at 25-30
Mpc. This indicates that blending could potentially be a substantial
source of error in the Cepheid distance scale.

As the result of blending with other unresolved stars, the Cepheids
appear brighter than they really are when observed in distant galaxies
with {\em HST}.  As we compare them with mostly unblended LMC
Cepheids, this leads to systematically low distances to galaxies
observed with {\em HST}, and therefore to systematically high
estimates of $H_0$.  The sign of the effect of blending on $H_0$
is opposite to that caused by the lower LMC distance (e.g. Udalski
2000; Stanek et al.~2000) and might be of comparable value, as
discussed in this paper. It should be stressed that blending is a
factor which contributes in only one direction, and therefore it will
not average out when a large sample of galaxies is considered.

A potential solution to the problem of blending could be the Cepheid
period - flux amplitude relation. Unfortunately this relation is not
universal, as Paczy\'nski \& Pindor (2000) find that SMC Cepheids have
lower amplitudes than those in the LMC and in our Galaxy. They suggest
metallicity as the most natural reason for this difference. If this is
the case and the metallicity dependence could be calibrated, the
Cepheid period - flux amplitude relation could offer the possibility
to obtain Cepheid distance determinations not affected by blending.

\acknowledgments{ Janusz Kaluzny and Bohdan Paczy\'nski have provided
us with helpful comments on the manuscript.  We would like to thank
Janusz Kaluzny for providing us with his database management codes,
Peter Stetson for the {\em HST} WFPC2 point-spread functions and Doug
Mink for assistance in automating image astrometry. This work was
partially based on observations with the NASA/ESA Hubble Space
Telescope, obtained from the data Archive at the Space Telescope
Science Institute, which is operated by the Association of
Universities for Research in Astronomy, Inc. under NASA contract
No. NAS5-26555. Support for this work was provided by NASA through
Grant AR-08741 from the Space Telescope Science Institute.  BJM was
supported by the Foundation for Polish Science stipend for young
scientists and the Polish KBN grants 2P03D003.17 to Janusz Kaluzny and
2P03D025.19. DDS acknowledges support from the Alfred P. Sloan
Foundation. Partial support for KZS was provided by NASA through
Hubble Fellowship Grant HF-01124.01-A from the Space Telescope Science
Institute.}

\begin{small}
\tablenum{1}
\begin{planotable}{crcccccccccc}
\tablewidth{39pc}
\tablecaption{\sc The M33 Cepheid Blending Catalog}
\tablehead{\colhead{Name} & \colhead{$P$} & \colhead{$\langle V \rangle$} &
\colhead{$\langle I \rangle$} & \colhead{$\langle B \rangle$} &
\colhead{$N_V$} & \colhead{$S_V$} & \colhead{$N_I$} & 
\colhead{$S_I$} & \colhead{$N_B$} & \colhead{$S_B$} & \colhead{$\sigma_V$}}
\startdata
 D33J013353.5+304744.1 &  4.31 &   21.17 & \nodata &   21.55 &       5 &    0.91 &       4 &    1.24 & \nodata & \nodata & 22.00\\ 
 D33J013359.4+304214.2 &  4.78 &   21.62 &   20.76 & \nodata &       3 &    0.26 &       1 &    0.07 & \nodata & \nodata & 21.13\\ 
 D33J013347.8+304627.4 &  4.86 &   20.83 & \nodata &   21.38 &       0 &    0.00 &       0 &    0.00 & \nodata & \nodata & 21.82\\ 
 D33J013357.0+304826.4 &  5.00 &   21.51 &   21.09 &   21.88 &       2 &    0.23 & \nodata & \nodata & \nodata & \nodata & 21.98\\ 
 D33J013400.4+304808.8 &  5.10 &   21.23 &   20.28 &   21.99 &       6 &    1.18 & \nodata & \nodata & \nodata & \nodata & 22.12\\ 
 D33J013408.6+303754.8 &  5.32 &   21.14 & \nodata &   21.70 &       2 &    0.24 &       2 &    0.34 &       1 &    0.07 & 21.26\\ 
 D33J013347.5+304456.2 &  5.35 &   99.00 & \nodata &   20.60 & \nodata & \nodata & \nodata & \nodata &       2 &    0.41 & 21.16\\ 
 D33J013421.6+304415.9 &  5.36 &   20.84 &   19.49 &   21.80 &       2 &    0.26 & \nodata & \nodata &       2 &    0.21 & 21.93\\ 
 D33J013405.5+304133.3 &  5.38 &   21.86 & \nodata &   22.43 &       0 &    0.00 &       2 &    0.24 & \nodata & \nodata & 21.18\\ 
 D33J013357.0+303117.5 &  5.55 &   20.60 &   19.52 &   21.19 &       1 &    0.19 &       2 &    0.56 &       0 &    0.00 & 22.15\\ 
 D33J013351.2+303001.0 &  5.60 &   20.50 &   19.57 &   21.15 &       1 &    0.10 &       1 &    0.24 & \nodata & \nodata & 22.13\\ 
 D33J013350.0+304346.7 &  5.64 &   21.02 &   20.59 &   21.38 & \nodata & \nodata & \nodata & \nodata &       2 &    0.37 & 21.29\\ 
 D33J013417.1+303932.9 &  5.64 &   19.88 &   19.41 &   20.26 &       3 &    0.94 & \nodata & \nodata & \nodata & \nodata & 21.88\\ 
 D33J013420.6+304244.2 &  5.70 &   21.32 &   20.39 &   22.00 &       2 &    0.27 & \nodata & \nodata &       2 &    0.26 & 22.10\\ 
 D33J013426.8+304357.7 &  5.70 &   19.83 &   19.46 & \nodata &       2 &    0.25 & \nodata & \nodata &       1 &    0.21 & 22.13\\ 
 D33J013350.6+304734.9 &  5.74 &   21.51 &   20.63 &   22.13 &       2 &    0.17 &       3 &    0.40 & \nodata & \nodata & 22.07\\ 
 D33J013405.0+303557.5 &  5.89 &   21.14 &   20.45 &   21.62 &       2 &    0.20 & \nodata & \nodata & \nodata & \nodata & 21.36\\ 
 D33J013407.9+303831.6 &  5.90 &   20.23 &   19.75 &   20.72 &       6 &    0.97 &       6 &    1.35 &       3 &    1.08 & 21.04\\ 
 D33J013424.9+304431.2 &  5.91 &   21.29 &   20.09 &   22.31 &       4 &    0.80 & \nodata & \nodata &       0 &    0.00 & 21.88\\ 
 D33J013346.9+304334.2 &  6.00 &   20.84 &   19.88 &   21.39 & \nodata & \nodata & \nodata & \nodata &       6 &    0.81 & 21.02\\ 
 D33J013349.6+304744.7 &  6.00 &   21.82 &   20.39 & \nodata &       1 &    0.21 &       1 &    0.30 & \nodata & \nodata & 22.17\\ 
 D33J013350.6+303445.8 &  6.03 &   21.03 &   20.32 &   21.69 &       6 &    1.18 &       3 &    0.82 & \nodata & \nodata & 21.21\\ 
 D33J013356.7+304838.6 &  6.12 &   20.90 &   20.09 &   21.32 &       1 &    0.07 & \nodata & \nodata & \nodata & \nodata & 21.94\\ 
 D33J013349.4+303009.4 &  6.78 &   21.03 &   20.33 &   21.19 &       4 &    0.33 &       3 &    0.25 & \nodata & \nodata & 22.12\\ 
 D33J013406.4+304003.7 &  6.93 &   20.47 &   19.48 &   21.06 &       2 &    0.20 &       2 &    0.29 & \nodata & \nodata & 20.89\\ 
 D33J013428.3+303900.4 &  6.99 &   21.15 &   20.29 &   21.65 &       1 &    0.08 &       1 &    0.12 & \nodata & \nodata & 22.47\\ 
 D33J013410.3+303934.8 &  7.09 &   21.39 &   20.88 &   21.98 &       1 &    0.07 & \nodata & \nodata & \nodata & \nodata & 21.28\\ 
 D33J013422.5+304408.4 &  7.66 &   21.23 &   20.14 &   22.01 &       3 &    0.35 & \nodata & \nodata &       1 &    0.23 & 21.99\\ 
 D33J013332.4+303143.3 &  7.97 &   21.40 &   20.22 &   22.59 &       1 &    0.35 & \nodata & \nodata &       0 &    0.00 & 21.87\\ 
 D33J013348.8+303415.8 &  7.97 &   20.26 & \nodata &   20.91 &       2 &    0.23 &       2 &    0.55 & \nodata & \nodata & 21.30\\ 
 D33J013356.2+303909.1 &  8.57 &   20.51 &   19.26 & \nodata &       3 &    0.71 &       4 &    1.27 &       3 &    0.97 & 20.39\\ 
 D33J013406.6+303816.8 &  8.58 &   20.21 & \nodata & \nodata &       2 &    0.13 &       1 &    0.09 &       1 &    0.13 & 20.97\\ 
 D33J013337.5+303305.1 &  8.98 &   20.97 &   19.99 & \nodata &       2 &    0.19 & \nodata & \nodata &       2 &    0.32 & 21.41\\ 
 D33J013333.5+303320.5 &  9.02 &   19.82 &   19.34 &   20.09 &       2 &    0.97 & \nodata & \nodata &       2 &    2.63 & 21.35\\ 
 D33J013346.3+302908.9 &  9.12 &   20.51 &   19.76 &   20.90 &       2 &    0.21 &       3 &    0.21 & \nodata & \nodata & 22.23\\ 
 D33J013352.7+303416.2 &  9.22 &   21.02 &   19.34 &   21.72 &       4 &    0.47 &       2 &    0.22 &       1 &    0.15 & 21.29\\ 
 D33J013350.8+304715.5 &  9.72 &   20.73 &   20.02 &   21.42 &       0 &    0.00 &       0 &    0.00 & \nodata & \nodata & 22.04\\ 
 D33J013421.1+304415.5 &  9.98 &   20.61 &   19.78 &   21.10 &       3 &    0.33 & \nodata & \nodata &       3 &    0.58 & 21.92\\ 
 D33J013408.8+303946.5 & 10.11 &   20.47 &   19.42 &   21.07 &       1 &    0.08 &       3 &    0.72 & \nodata & \nodata & 21.12\\ 
 D33J013355.0+303537.0 & 10.13 &   20.50 &   19.60 &   21.07 &       1 &    0.07 &       0 &    0.00 &       1 &    0.18 & 21.15\\ 
\enddata
\label{tab:cep}
\end{planotable}
\end{small}

\begin{small}
\tablenum{1}
\begin{planotable}{crcccccccccc}
\tablewidth{39pc}
\tablecaption{\sc Continued.}
\tablehead{\colhead{Name} & \colhead{$P$} & \colhead{$\langle V \rangle$} &
\colhead{$\langle I \rangle$} & \colhead{$\langle B \rangle$} &
\colhead{$N_V$} & \colhead{$S_V$} & \colhead{$N_I$} &
\colhead{$S_I$} & \colhead{$N_B$} & \colhead{$S_B$} & \colhead{$\sigma_V$}}
\startdata
 D33J013342.1+303210.7 & 10.38 &   21.01 &   19.14 & \nodata &       2 &    0.17 & \nodata & \nodata &       2 &    0.50 & 21.40\\ 
 D33J013356.1+303903.0 & 10.42 &   20.43 & \nodata & \nodata &       2 &    0.18 &       5 &    0.54 &       0 &    0.00 & 20.41\\ 
 D33J013335.6+303649.2 & 10.70 &   20.73 &   20.24 &   22.10 &       1 &    0.07 &       1 &    0.09 & \nodata & \nodata & 21.22\\ 
 D33J013327.4+303550.9 & 11.21 &   20.49 &   19.79 &   21.33 &       2 &    0.16 &       2 &    0.21 &       1 &    0.14 & 21.64\\ 
 D33J013325.7+303426.6 & 11.45 &   20.19 &   19.47 &   20.82 &       0 &    0.00 & \nodata & \nodata &       0 &    0.00 & 21.93\\ 
 D33J013353.4+303308.5 & 11.48 &   21.38 &   20.63 & \nodata &       2 &    0.13 &       1 &    0.08 &       0 &    0.00 & 21.48\\ 
 D33J013335.5+303330.2 & 11.52 &   20.55 &   19.69 &   21.36 &       1 &    0.14 & \nodata & \nodata &       1 &    0.24 & 21.40\\ 
 D33J013357.4+304113.9 & 11.62 &   20.25 &   19.23 &   20.78 &       3 &    1.39 &       4 &    1.35 & \nodata & \nodata & 20.94\\ 
 D33J013337.7+303218.9 & 11.88 &   20.80 &   19.74 & \nodata &       0 &    0.00 & \nodata & \nodata &       0 &    0.00 & 21.40\\ 
 D33J013420.3+304351.9 & 11.97 &   20.33 &   19.43 &   21.08 &       1 &    0.10 & \nodata & \nodata &       1 &    0.14 & 21.97\\ 
 D33J013357.6+303805.4 & 12.34 &   20.48 &   19.71 & \nodata &       1 &    0.16 &       1 &    0.08 & \nodata & \nodata & 20.92\\ 
 D33J013351.1+304400.4 & 12.35 &   21.06 &   20.05 &   22.01 & \nodata & \nodata & \nodata & \nodata &       0 &    0.00 & 21.39\\ 
 D33J013346.6+304821.8 & 12.36 &   20.13 &   18.38 &   21.28 &       1 &    0.07 &       1 &    0.08 & \nodata & \nodata & 21.97\\ 
 D33J013256.3+303437.1 & 12.82 &   20.57 &   19.72 & \nodata &       0 &    0.00 & \nodata & \nodata &       0 &    0.00 & 22.66\\ 
 D33J013349.8+303758.7 & 12.91 &   20.16 &   19.74 & \nodata &       1 &    0.06 &       2 &    0.20 & \nodata & \nodata & 20.70\\ 
 D33J013350.0+303014.9 & 12.98 &   20.64 &   19.32 &   21.41 &       1 &    0.07 &       2 &    0.48 & \nodata & \nodata & 22.09\\ 
 D33J013402.8+304145.7 & 13.04 &   19.93 &   18.80 &   20.60 &       2 &    0.18 &       4 &    0.75 & \nodata & \nodata & 21.13\\ 
 D33J013345.9+304421.4 & 13.12 &   20.48 & \nodata &   21.51 & \nodata & \nodata & \nodata & \nodata &       1 &    0.42 & 21.22\\ 
 D33J013331.7+303931.1 & 13.17 &   20.13 &   19.37 &   20.79 &       2 &    0.19 &       1 &    0.14 &       1 &    0.32 & 21.55\\ 
 D33J013325.5+304037.5 & 13.30 &   20.25 &   19.30 &   21.38 &       1 &    0.06 &       1 &    0.06 & \nodata & \nodata & 21.86\\ 
 D33J013408.1+303931.9 & 13.32 &   20.32 &   19.73 &   20.55 &       5 &    0.78 &       3 &    0.38 & \nodata & \nodata & 21.07\\ 
 D33J013351.2+303758.2 & 13.56 &   19.80 & \nodata & \nodata &       0 &    0.00 &       0 &    0.00 & \nodata & \nodata & 20.56\\ 
 D33J013402.5+303628.0 & 13.65 &   20.36 &   19.43 &   20.77 &       3 &    0.39 & \nodata & \nodata & \nodata & \nodata & 20.93\\ 
 D33J013408.4+303817.2 & 14.35 &   20.17 &   19.14 & \nodata &       2 &    0.36 &       2 &    0.39 &       2 &    0.85 & 21.14\\ 
 D33J013405.9+303928.9 & 14.60 &   19.92 &   19.21 & \nodata &       2 &    0.38 &       2 &    0.25 & \nodata & \nodata & 20.90\\ 
 D33J013327.8+303423.2 & 14.85 &   19.51 &   18.81 &   19.93 &       2 &    0.15 & \nodata & \nodata &       1 &    0.10 & 21.86\\ 
 D33J013401.3+304026.9 & 14.85 &   19.86 &   19.26 &   20.17 &       4 &    1.48 &       4 &    0.70 & \nodata & \nodata & 20.84\\ 
 D33J013334.4+303530.2 & 15.84 &   19.76 &   18.82 &   20.24 &       2 &    0.32 &       2 &    0.20 &       3 &    0.89 & 21.20\\ 
 D33J013403.9+303615.8 & 16.28 &   20.26 &   19.22 &   21.00 &       0 &    0.00 & \nodata & \nodata & \nodata & \nodata & 21.09\\ 
 D33J013353.4+303535.3 & 17.48 &   19.68 &   18.94 & \nodata &       1 &    0.11 &       1 &    0.17 & \nodata & \nodata & 21.05\\ 
 D33J013330.2+303637.4 & 17.98 &   20.17 &   18.94 &   20.61 &       0 &    0.00 &       0 &    0.00 &       1 &    0.08 & 21.37\\ 
 D33J013346.6+304645.9 & 18.82 &   19.70 &   18.84 &   20.60 &       0 &    0.00 &       0 &    0.00 & \nodata & \nodata & 21.88\\ 
 D33J013406.8+303940.2 & 18.89 &   19.76 &   18.86 &   20.51 &       5 &    0.46 &       4 &    0.34 & \nodata & \nodata & 20.91\\ 
 D33J013330.2+303803.7 & 19.92 &   19.99 &   19.19 &   20.59 &       2 &    0.39 &       2 &    0.16 &       2 &    0.28 & 21.50\\ 
 D33J013326.2+303319.4 & 19.99 &   20.16 &   19.19 &   21.03 &       0 &    0.00 & \nodata & \nodata &       0 &    0.00 & 22.01\\ 
 D33J013331.5+303351.2 & 20.17 &   19.09 & \nodata &   20.49 &       0 &    0.00 & \nodata & \nodata &       0 &    0.00 & 21.46\\ 
 D33J013343.4+304356.5 & 20.19 &   19.33 &   18.83 &   19.51 & \nodata & \nodata & \nodata & \nodata &       4 &    1.70 & 21.40\\ 
 D33J013419.0+304441.4 & 20.21 &   20.17 &   18.79 &   21.27 &       2 &    0.24 & \nodata & \nodata &       0 &    0.00 & 21.74\\ 
 D33J013255.6+303512.9 & 21.06 &   20.02 &   19.16 &   20.99 &       0 &    0.00 & \nodata & \nodata &       0 &    0.00 & 22.57\\ 
 D33J013401.7+303923.1 & 21.68 &   19.81 &   18.74 & \nodata &       1 &    0.08 &       0 &    0.00 &       3 &    0.53 & 20.84\\ 
\enddata
\end{planotable}
\end{small}

\begin{small}
\tablenum{1}
\begin{planotable}{crcccccccccc}
\tablewidth{39pc}
\tablecaption{\sc Continued.}
\tablehead{\colhead{Name} & \colhead{$P$} & \colhead{$\langle V \rangle$} &
\colhead{$\langle I \rangle$} & \colhead{$\langle B \rangle$} &
\colhead{$N_V$} & \colhead{$S_V$} & \colhead{$N_I$} &
\colhead{$S_I$} & \colhead{$N_B$} & \colhead{$S_B$} & \colhead{$\sigma_V$}}
\startdata 
 D33J013351.4+303830.7 & 21.79 &   20.20 &   18.87 & \nodata &       1 &    0.10 &       4 &    0.73 &       0 &    0.00 & 20.32\\ 
 D33J013330.4+303555.0 & 22.09 &   19.35 &   18.66 &   20.28 &       0 &    0.00 &       0 &    0.00 &       0 &    0.00 & 21.41\\ 
 D33J013333.3+303747.1 & 22.68 &   20.51 &   19.30 & \nodata &       0 &    0.00 &       1 &    0.40 &       0 &    0.00 & 21.43\\ 
 D33J013417.6+303819.7 & 23.30 &   20.22 &   19.14 &   21.12 &       1 &    0.11 & \nodata & \nodata &       0 &    0.00 & 21.66\\ 
 D33J013350.7+303544.2 & 23.31 &   19.88 &   18.89 & \nodata &       0 &    0.00 &       1 &    0.06 & \nodata & \nodata & 21.23\\ 
 D33J013358.8+303719.7 & 24.56 &   20.01 &   18.81 & \nodata &       1 &    0.06 & \nodata & \nodata & \nodata & \nodata & 21.14\\ 
 D33J013350.6+304754.7 & 26.48 &   19.93 &   18.62 &   21.10 &       1 &    0.11 &       1 &    0.36 & \nodata & \nodata & 22.21\\ 
 D33J013354.3+304111.2 & 27.98 &   19.08 & \nodata &   19.92 &       2 &    0.15 &       1 &    0.10 & \nodata & \nodata & 20.83\\ 
 D33J013332.9+303548.4 & 30.29 &   19.44 &   18.50 & \nodata &       3 &    0.22 &       3 &    0.29 &       3 &    0.40 & 21.15\\ 
 D33J013302.3+303632.9 & 30.53 &   19.52 &   18.65 &   20.44 &       0 &    0.00 & \nodata & \nodata & \nodata & \nodata & 22.78\\ 
 D33J013329.5+303556.9 & 30.66 &   19.58 &   18.65 &   20.55 &       1 &    0.10 &       1 &    0.06 &       1 &    0.15 & 21.43\\ 
 D33J013354.8+304106.5 & 33.95 &   19.14 &   18.11 &   20.00 &       1 &    0.07 &       2 &    0.28 &       2 &    0.22 & 20.79\\ 
 D33J013352.4+303844.2 & 35.94 &   19.11 &   18.10 & \nodata &       0 &    0.00 &       1 &    0.10 &       0 &    0.00 & 20.20\\ 
 D33J013327.5+303707.2 & 37.28 &   19.11 &   18.15 &   20.05 &       0 &    0.00 &       0 &    0.00 &       0 &    0.00 & 21.50\\ 
 D33J013350.9+303336.1 & 37.57 &   20.48 &   19.03 &   21.74 &       0 &    0.00 &       0 &    0.00 & \nodata & \nodata & 21.40\\ 
 D33J013329.3+303744.4 & 46.01 &   19.10 &   18.07 &   20.36 &       0 &    0.00 &       0 &    0.00 &       0 &    0.00 & 21.47\\ 
 D33J013359.4+303226.7 & 50.28 &   19.87 &   18.45 &   20.82 &       0 &    0.00 & \nodata & \nodata & \nodata & \nodata & 21.75\\ 
 D33J013347.5+304423.2 & 55.98 &   19.46 &   18.28 &   20.71 & \nodata & \nodata & \nodata & \nodata &       0 &    0.00 & 21.12\\ 
 D33J013403.8+303911.1 & 57.44 &   19.56 &   18.21 &   20.64 &       0 &    0.00 &       0 &    0.00 &       1 &    0.20 & 20.90\\ 
 D33J013337.5+303138.5 & 57.63 &   19.17 &   18.03 & \nodata &       1 &    0.07 & \nodata & \nodata &       1 &    0.36 & 21.48\\ 
 D33J013351.3+303900.9 & 62.00 &   17.44 & \nodata & \nodata &       0 &    0.00 &       0 &    0.00 &       2 &    0.15 & 19.85\\ 
 D33J013351.8+303951.0 & 67.32 &   18.32 &   17.54 & \nodata &       0 &    0.00 &       1 &    0.07 &       1 &    0.10 & 19.73\\ 
\enddata
\end{planotable}
\end{small}

\end{document}